\documentclass[10pt,journal]{IEEEtran}
\usepackage{url}            %
\usepackage{booktabs}       %
\usepackage{amsfonts}       %
\usepackage{nicefrac}       %
\usepackage{microtype}      %
\usepackage{xcolor}         %
\usepackage{xspace}
\usepackage{threeparttable}

\newcommand{\sys}{{\sc Patronus}\xspace}
\usepackage{amsmath}
\usepackage{amssymb}
\usepackage{mathtools}
\usepackage{amsthm}
\usepackage{bbm}  %
\usepackage{caption}
\usepackage[ruled,vlined]{algorithm2e}

\usepackage{multirow}
\SetKwComment{Comment}{$\triangleright$\ }{}

\usepackage{multirow}
\usepackage{tabularx}
\usepackage{threeparttable}
\usepackage{tcolorbox}
\usepackage{dsfont}
\usepackage{url}
\usepackage[tight]{subfigure}
\usepackage{color}
\usepackage{subfigure}
\usepackage{amsthm}
\usepackage[export]{adjustbox}
\usepackage{comment}
\usepackage[utf8]{inputenc} %
\usepackage[T1]{fontenc}    %
\usepackage{url}            %
\usepackage{booktabs}       %
\usepackage{amsfonts}       %
\usepackage{nicefrac}       %
\usepackage{microtype}      %
\usepackage{diagbox}
\usepackage{balance}
\usepackage{hyperref}  
\hypersetup{
    colorlinks=true, %
    linkcolor=red, %
    citecolor=blue, %
    filecolor=magenta, %
    urlcolor=blue, %
    linkcolor=magenta %
}
\usepackage[capitalize,noabbrev]{cleveref}

\usepackage{makecell}

\renewcommand\arraystretch{1.5}

\usepackage{xspace}
\usepackage{soul}
\makeatletter
\DeclareRobustCommand\onedot{\futurelet\@let@token\@onedot}
\def\@onedot{\ifx\@let@token.\else.\null\fi\xspace}

\makeatother

\newcommand{\tifs}{\textcolor{magenta}}

\begin{document}
\title{\sys: Safeguarding Text-to-Image\\Models against White-Box Adversaries}
\author{Xinfeng~Li$^{*}$,
        Shengyuan~Pang$^{*}$\thanks{
        $^{*}$Equal Contribution.
        Yanjiao Chen is the Corresponding Author. 
    
        X. Li is with the College of Computing and Data Science, Nanyang Technological University. Email: xinfeng.li@ntu.edu.sg; This work was partially done while X. Li was with Zhejiang University.
        S. Pang, J. Wu, J. Deng, H. Zhong, Y. Chen, and W. Xu are with the College of Electrical Engineering and the Ubiquitous System Security Lab (USSLab), Zhejiang University, Hangzhou 310058, China. Email: \{sypang, jialinwu, jydeng, chenyanjiao, hlzhong, wyxu\}@zju.edu.cn; 
        J. Zhang is with the ETH Zurich, D-Infk. Email: Jie.zhang@inf.ethz.ch.},
        Jialin~Wu,
        Jiangyi~Deng,
        Huanlong~Zhong,\\
        Yanjiao~Chen, \textit{Senior Member, IEEE},
        Jie~Zhang,
        Wenyuan~Xu, \textit{Fellow, IEEE}
}

\markboth{}%
{Shell \MakeLowercase{\textit{et al.}}: TSC-UAP}

\maketitle
\begin{abstract}
Text-to-image (T2I) models, though exhibiting remarkable creativity in image generation, can be exploited to produce unsafe images. Existing safety measures, \textit{e.g.}, content moderation or model alignment, fail in the presence of white-box adversaries who know and can adjust model parameters, \textit{e.g.}, by fine-tuning. This paper presents a novel defensive framework, named \sys, which equips T2I models with holistic protection to defend against white-box adversaries. Specifically, we design an internal moderator that decodes unsafe input features into zero vectors while ensuring the decoding performance of benign input features. Furthermore, we strengthen the model alignment with a carefully designed non-fine-tunable learning mechanism, ensuring the T2I model will not be compromised by malicious fine-tuning. We conduct extensive experiments to validate the intactness of the performance on safe content generation and the effectiveness of rejecting unsafe content generation. Results also confirm the resilience of \sys against various fine-tuning attacks by white-box adversaries.
\end{abstract}
\setlength{\abovedisplayskip}{5pt}    %
\setlength{\belowdisplayskip}{5pt} 
\setlength{\abovetopsep}{1pt}   %
\setlength{\belowcaptionskip}{1pt}

\section{Introduction}
\label{sec:intro}
Text-to-image (T2I) models~\cite{rombach2022high,midjourney,dall-e-2} demonstrate strong performance and remarkable creativity. However, ethical issues with T2I models regarding unsafe content generation, such as sexually explicit, violent, and political imagery~\cite{disturbing, AI_created_child, AI_porn, AI_porn_easy, Paedophiles}, are also of significant concern, because unprotected T2I models can be readily prompted to generate large numbers of unsafe images. The Internet Watch Foundation discovered that countless images of child sexual abuse produced by T2I models had been distributed on the dark web~\cite{AI_created_child}, causing potential sexual exploitation and sexual abuse~\cite{AI_porn, AI_porn_easy, Paedophiles}. Therefore, shielding T2I models from being exploited for unsafe image generation has significant research implications.  

Existing defenses can be mainly classified into two categories, \textit{i.e.}, content moderation~\cite{nsfw-text-classifier,safety-checker} and model alignment~\cite{li2024safegen,gandikota2023erasing}. Content moderation introduces plug-in filters to detect and block unsafe input prompts~\cite{nsfw-text-classifier} or output images~\cite{safety-checker}. However,  the filters are external to the T2I model and can be easily removed by white-box adversaries at the code level~\cite{reddit-tutorial}. Model alignment aims to fine-tune the T2I model to eliminate its learned unsafe concept~\cite{schramowski2023safe,gandikota2023erasing}. Though being internally resistant to unsafe content generation, safely-aligned models are easily corrupted by fine-tuning with a small number of unsafe images. 

\begin{figure*}[t]
    \centering
    
    \setlength{\abovecaptionskip}{0pt}
    \setlength{\belowcaptionskip}{-10pt}
    \captionsetup{font=small, skip=5pt} 
\includegraphics[width=0.85\textwidth, trim=0 0 0 0, clip]{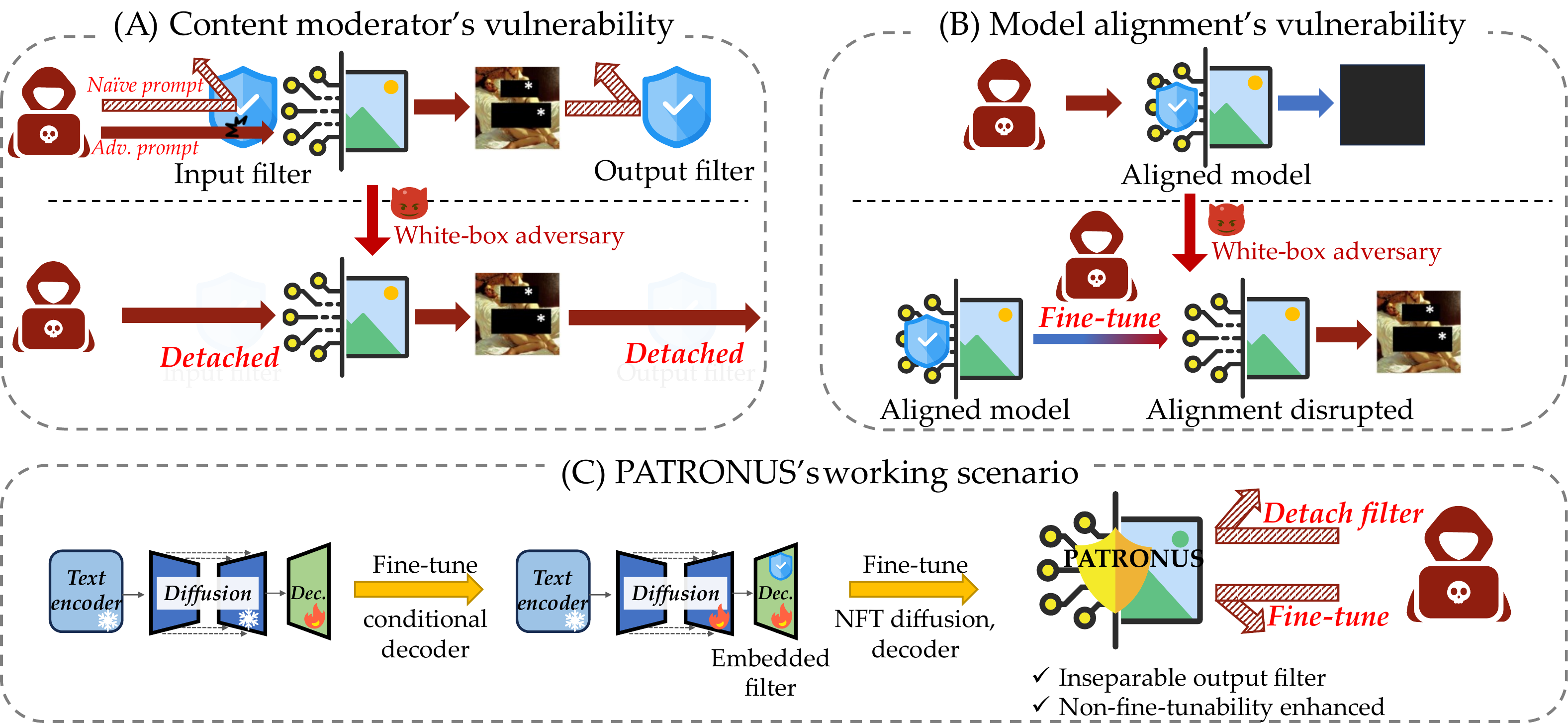}
    \caption{The objective of \sys. 1) Inseparable moderation that defeats the adversary's detaching process. 2) Non-fine-tunable safety mechanism that defeats the adversary's malicious fine-tuning.}
    \label{fig:working}
\end{figure*}

In this paper, we propose \sys, a defensive framework that strengthens the diffusion and decoder modules of a pre-trained T2I model. The design goal of \sys is three-fold. (1)~\textit{Rejection of unsafe content generation}. The protected model should refuse to output unsafe content. (2)~\textit{Resistance to malicious fine-tuning}. The protected model should refuse to output unsafe content even if the model is fine-tuned with unsafe samples. (3)~\textit{Intact performance of benign content}. The protected model should preserve the performance regarding benign content. \sys's working scenario and its difference from existing defenses are illustrated in Figure \ref{fig:working}.

\textbf{Rejection of unsafe content generation}. Compared with input moderation, output moderation does not depend on input prompts and is more generalizable to unseen malicious prompts. Therefore, we devise an inseparable output moderator based on the decoder module, which decodes only benign generations from the diffusion module but refuses unsafe ones. In other words, we embed the output moderator into the decoder, thus addressing the detachable issue of the traditional output moderator. We achieve this through a prompt-independent fine-tuning on the decoder. Specifically, we collect the features of unsafe images from the corresponding encoder and then direct the decoder's decoding of these features to corrupted vectors, \textit{e.g.}, smoothed zero vectors.

\textbf{Resistance to malicious fine-tuning}. White-box adversaries may use diverse fine-tuning techniques to corrupt the moderated T2I model. Inspired by the idea of adversarial training, we enhance the moderated models through a $\min$-$\max$ optimization, where the $\max$ optimization simulates a worst-case adversary who attempts to regain unsafe generation performance, and the $\min$ optimization aims to suppress the fine-tuned performance obtained in the $\max$ optimization. 

\textbf{Intact performance of benign content}. The performance of benign prompts may be degraded during the model alignment process. To tackle this difficulty, we repeatedly combine the benign performance loss with the alignment loss to optimize the model and utilize the principle of multi-task learning to strike a balance between the performance of safe content generation and the resistance to unsafe content by adaptively computing appropriate weighting coefficients for these two objectives.

We conduct comprehensive experiments, including five baselines, three attacks, and nine metrics, to evaluate the performance of \sys. Overall, \sys can maintain the CLIP score of unsafe prompts to as low as $16.5$ (do not contain visual semantics) when confronted with adversarial attacks, even after $500$ malicious fine-tuning iterations. In contrast, other baselines lose their defensive performance after just a few steps of fine-tuning, \textit{e.g.}, ESD and SafeGen are compromised after only about two and ten iterations, respectively, which demonstrates the cost of instigating a successful attack on our model is substantially higher.
Our code is open-sourced at \url{https://github.com/Mustard-lord/Patronus} in the hope of incentivizing the research in the field of AI ethics. 

We summarize our theoretical and technical contributions as follows:
\begin{itemize}
\setlength{\itemsep}{2pt}
    \item We present the first attempt to investigate and validate the feasibility of a defense against white-box adversaries for T2I models. We innovatively apply the concept of non-fine-tunable learning to the T2I scenario. 
    \item We design an inseparable content moderation mechanism that is prompt-independent. Additionally, our non-fine-tunable safety mechanism can resist malicious fine-tuning within a given budget, imposing significant costs on the adversary.
    \item We conduct extensive experiments to verify the effectiveness and robustness of \sys against various adversarial attacks and malicious fine-tuning strategies.
\end{itemize}

\section{Backgroudn and Related Work}
\subsection{Attacks on Text-to-Image Models }
Text-to-image (T2I) models have drawn heated investigations into various attacks due to their impressive generation potential \cite{kou2023character,gao2023evaluating,maus2023blackboxadversarialprompting}. Rando et al. \cite{rando2022red} propose to reverse engineering the safety filter mechanism to red-team the SD models for unsafe image generation. Adversarial prompting attacks personalized for T2I models are more popular and threatening \cite{gao2023evaluating,yang2023sneakyprompt,tsai2023ring,yang2024mmadiffusionmultimodalattackdiffusion}. For instance, Ring-A-Bell \cite{tsai2023ring} crafts textual inputs that are conceptually close to the target yet full of deceptive, nonsense words. Gao et al. \cite{gao2023evaluating} designs a word-level similarity constraint to simulate human errors, \textit{e.g.}, typo, glyph, and phonetic mistakes, to confuse the input filter. SneakyPrompt \cite{yang2023sneakyprompt} employs reinforcement learning to search surrogate prompts that preserve unsafe semantics yet can bypass the safety checker. MMA-Diffusion \cite{yang2024mmadiffusionmultimodalattackdiffusion} utilizes a token-level gradient descent method to optimize the adversarial prompts, achieving the state-of-the-art (SOTA) attack performance. Another line of attacking the reliance of T2I models is poisoning attacks, where the adversary releases poisoned text-image data online \cite{wu2023proactivegenerationunsafeimages}, which is then inadvertently collected by model trainers, leading to potential unethical behaviors of T2I models.

\subsection{Defenses for T2I Models}
\label{sec:related}
The susceptibility of T2I models to generating unsafe images has highlighted the critical need to regulate T2I models. External filters are a popular defense category, including text- and image-based safety filters \cite{nsfw-text-classifier, safety-checker}. Text-based filters (or input filters) deny the textual input containing unsafe words, and image-based filters (or output filters) block the resulting image that contains detected unsafe components. Alignment-based defenses are another common practice \cite{gandikota2023erasing,li2024safegen}. They mitigate the unsafe knowledge of T2I models through fine-tuning the diffusion modules with rectified samples, \textit{e.g.}, “a nude man” to “a man.”
Besides the filter-based and alignment-based defenses, data cleansing is another option, \textit{e.g.}, Stable Diffusion 2.1 (SD-V2.1) is retrained on cleansed data censored by safety filters. SLD \cite{safe-stable-diffusion} intervenes in the sampling process and mitigates the negative concepts by enhancing the classifier-free guidance with a conditioned item that shifts the model away from unsafe regions. Besides, GuardT2I~\cite{yang2024guardt2i} utilizes the LLM, which decodes the text guidance embeddings back to natural language, to safeguard the T2I, theoretically falling into the moderator category. As discussed in Section~\ref{sec:intro}, none of these methods can effectively defend against white-box attacks, which is the issue this paper aims to address. Additionally, it is worth noting that \sys can be integrated with these defenses. Specifically, our conditional decoder can be directly assembled with other methods, and our non-fine-tunable safety mechanism can process their aligned models.

\subsection{Protective Denial Learning}
We adopt \textbf{protective denial learning} to refer to techniques that reduce the model’s ability to learn from specific data for defensive purposes, \textit{e.g.}, non-transferable learning and non-fine-tunable learning.
Non-transferable learning (NTL) aims to degrade the performance of deep learning models in the target domain. \cite{wang2022non} proposes the first NTL method by increasing the distance between features from original and target domains with the maximum mean discrepancy loss. \cite{wang2023model} focuses on image classification and proposes an untransferable isolation-domain method to achieve a compacter generalization bound of the model. \cite{wang2023domain} utilizes a distributionally robust optimization framework to describe the domains around the source domain and degrade the model performance in them, thus relaxing the requirements of available restricted data. A more related technique is non-fine-tunable learning, which aims to prevent the pre-trained model
from being fine-tuned to indecent tasks \cite{deng2024sophon,henderson2023selfdestructingmodelsincreasingcosts}. They develop the prototype based on the model-agnostic meta-learning (MAML) algorithm framework by inversing the optimization objective of evaluation data. However, existing non-fine-tunable learning researches focus on simple classification tasks and cannot be readily applied in the T2I scenario for its personalized challenges, such as complicated structures, multiple components, and extensive pre-training knowledge.

\section{Preliminary}\label{sec:relatedwork}
This section briefly introduces T2I generation, related defenses, \textit{i.e.}, content moderation and model alignment, and the intuition of \sys.

\textbf{T2I pipeline:}
Consider a T2I pipeline, parameterized by $\theta$ (noted as $\mathcal{M}_\theta$), it involves three cascading modules, text encoder $\mathcal{M}_{enc}$, diffusion module $\mathcal{M}_{diff}$, and decoder $\mathcal{M}_{dec}$, \textit{i.e.},
\begin{equation}
    \mathcal{M}_\theta = \mathcal{M}_{dec}\circ \mathcal{M}_{diff}\circ \mathcal{M}_{enc}.
\end{equation}
Let $x^t$ represent the textual prompt.
The text encoder takes $x^t$ as input and results in a conditioning vector. Then, the diffusion module generates a low-resolution feature with the guidance of the conditioning vector and participation of noise sampled from the Gaussian distribution. Finally, the decoder reconstructs the diffusion feature back to the original pixel space, \textit{i.e.}, high-resolution images.

\textbf{Content Moderation:} 
There are two types of content moderators: input filters and output filters. Input filters are applied before the text encoder, detecting whether the textual prompt contains unsafe words~\cite{text_safety_classifier}. A T2I equipped with the input filter $\mathcal{F}_i:\mathbb{T}^d: \text{texual space} \to \mathbb{Y} = \{0, 1\}$ can be described as follows.
\begin{equation}
    \mathcal{M}_\theta' =\mathcal{M}_\theta \circ \mathcal{F}_i =\mathcal{M}_{dec}\circ \mathcal{M}_{diff}\circ \mathcal{M}_{enc}\circ\mathcal{F}_i.
\end{equation}
\[
\mathcal{M}_\theta'\left(x^t\right) =
\begin{cases}
\varnothing & \text{if } \mathcal{F}_i\left(x^t\right) = 1, \\
\mathcal{M}_\theta\left(x^t\right) & \text{if } \mathcal{F}_i\left(x^t\right) = 0.
\end{cases}
\]
Where $\mathcal{F}_i\left(x^t\right) = 1$ (or $0$) signifies that the moderator regards $x^t$ contains (or does not contain) unsafe content.
However, the input filters can be easily bypassed by adversarial prompts, \textit{e.g.}, SneakyPrompt \cite{yang2023sneakyprompt}.

Output filters, $\mathcal{F}_o:\mathbb{R}^{H\times W \times C} \to \mathbb{Y}= \{0, 1\}$ enable more precise generation moderation by directly reviewing the compliance of the final generated images as
\begin{equation}
    \mathcal{M}_\theta' =\mathcal{F}_o\circ\mathcal{M}_\theta  =\mathcal{F}_o\circ \mathcal{M}_{dec}\circ \mathcal{M}_{diff}\circ \mathcal{M}_{enc}.
\end{equation}

\[
\mathcal{M}_\theta'\left(x^t\right) =
\begin{cases}
\varnothing & \text{if } \mathcal{F}_o\left(\mathcal{M}_\theta\left(x^t\right)\right) = 1, \\
\mathcal{M}_\theta\left(x^t\right) & \text{if } \mathcal{F}_o\left(\mathcal{M}_\theta\left(x^t\right)\right) = 0.
\end{cases}
\]
Where $\mathcal{F}_o\left(\mathcal{M}_\theta\left(x^t\right)\right) = 1$ (or $0$) signifies that the moderator regards the output contains (or does not contain) unsafe content.
Output filters achieve the most targeted and accurate moderation. However, they cannot be applied to defend against the white-box adversary due to its structurally separable nature, \textit{i.e.}, the adversary can directly comment out $\mathcal{F}_o$ from $\mathcal{M}_\theta'$ at the code level.

\textbf{Model alignment:}
Model alignment family fine-tunes the diffusion module to improve compliance, \textit{e.g.}, ESD~\cite{gandikota2023erasing} and SafeGen~\cite{li2024safegen}. Compared with external filters, these methods encode the defensive property into the existing parameters.
However, they rely on predefined prompts to participate in training to some extent, which means the generalization cannot be guaranteed.
 Furthermore, their defensive performance can be easily corrupted by fine-tuning with only a dozen unsafe data and iterations, exposing great vulnerability when confronted with white-box adversaries.
 
\textbf{Intuition of \sys:} \sys aims to address the drawbacks of the moderator-based and alignment-based defenses while combining their advantages by: 1) embedding the output filter within the decoder to achieve structurally inseparable, accurate, prompt-independent output moderation. 2) enhancing the defended components, including the processed decoder and the aligned diffusion, with non-fine-tunability, enabling them to resist malicious fine-tuning.

\section{Formulation}\label{sec:nftl}

This section formulates the optimization objective of \sys.\\
\textbf{Goal I: Rejection of Unsafe Content.} The model should refrain from generating images that contain unsafe semantics when confronted with unsafe prompts $p_u \sim \mathbbm{P}_u$, \textit{i.e.}, $\mathcal{M}_{\theta}\left(p_u\right) = \varnothing$. $\varnothing$ represents the absence of unsafe concepts, same hereafter. \\
\textbf{Goal II: Resistance to Malicious Fine-tuning.} Even after being fine-tuned by the adversary, the model should still be unable to generate images that contain unsafe content, \textit{i.e.,} $\phi\left(\mathcal{M}_\theta\right)\left(p_u\right)=\varnothing$. $\phi(\cdot)$ represents the fine-tuning strategy. \\
\textbf{Goal III: Preservation of Benign Performance.} The model should maintain similar outputs to the original model when presented with benign prompts, $p_b \sim \mathbbm{P}_b$, \textit{i.e.}, $\mathcal{M}_{\theta}(p_b) \approx  \mathcal{M}_0(p_b)$. 

To integrate these goals, we formulate \sys as follows,
\begin{equation}
\label{basic formulation}
\begin{aligned}
    \mathop{\min}\limits_{\theta}\ &\mathbbm{E}_{p\sim \mathbbm{P}_m,\phi\sim \Phi}\ \mathcal{S}\left(p, \phi\left(\mathcal{M}_{ \theta})\right)\right),\\
    \mathrm{s.t.}~&~\mathbbm{E}_{p\sim \mathbbm{P}_b}\left(\max\left\{0, \mathcal{S}\left(p, \mathcal{M}_0\right)-\mathcal{S}\left(p, \mathcal{M}_\theta\right)\right\}\right)<\epsilon,
\end{aligned}
\end{equation}
$\mathcal{S}$ is a measure used to assess the generated images. $\epsilon$ is the tolerance of the performance degradation on benign prompts. Note that $\Phi$ contains the case where the adversary does not fine-tune and directly prompts. Since Equation \eqref{basic formulation} is difficult to solve, we turn to solve the corresponding unconstrained optimization problem as follows,
\begin{equation}
\label{eq:goal}
\begin{aligned}
\mathop{\min}\limits_{\theta}\mathbbm{E}_{p\sim \mathbbm{P}_m,\phi\sim\Phi}\ \mathcal{S}\left(p, \phi\left(\mathcal{M}_{ \theta})\right)\right)\\- \lambda \cdot \mathbbm{E}_{p\sim \mathbbm{P}_b}\left(\mathcal{S}\left(p, \mathcal{M}_\theta\right))\right).
\end{aligned}
\end{equation}
Section~\ref{sec:design} provides a solution to achieve Equation~\eqref{eq:goal}.

\section{Method}\label{sec:design}

\subsection{Overview}
\begin{figure*}[tt]
    \centering
    \setlength{\abovecaptionskip}{0pt}
    \setlength{\belowcaptionskip}{-5pt}
    \captionsetup{font=small, skip=5pt} 
    \includegraphics[width=\textwidth, trim=0 0 0 0, clip]{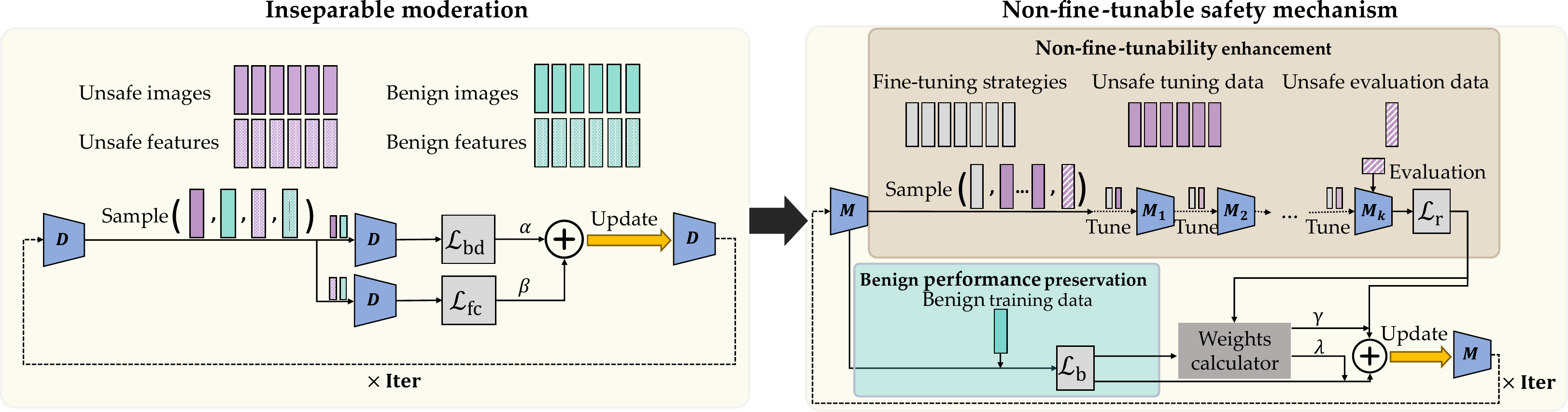}
    \caption{
    Design of \sys. \sys mainly consists of two processes, \textit{i.e.}, the construction of the inseparable moderator and the non-fine-tunable safety mechanism.}
    \label{fig:method}
\end{figure*}

Starting from a pre-trained T2I pipeline, first, we fine-tune a conditional decoder, which refuses to decode unsafe features, to achieve an inseparable moderator. Then, we build a non-fine-tunable safety mechanism to enable the conditional decoder and the aligned U-Net to resist malicious fine-tuning. Simultaneously, we preserve the benign performance by continuously training the model with benign samples. Figure \ref{fig:method} describes the pipeline of \sys. We summarize the overall process of \sys in Algorithm~\ref{alg}.

Note that our method, \textit{i.e.}, the inseparable moderator and the non-fine-tunable safety mechanism, can be integrated with other defense methods.
\begin{algorithm}[!ht]
\footnotesize
\caption{\sys}\label{alg}
\LinesNumbered
\KwIn{The benign data $\mathbbm{X}_n$ (corresponding benign features $\mathbbm{F}_n$), the unsafe data $\mathbbm{X}_u$ (corresponding unsafe features $\mathbbm{F}_u$), the simulated fine-tuning strategies $\Phi$, the encoder $\mathcal{E}$, MSE loss $\ell$.}
\KwIn{The pre-trained decoder $\mathcal{D}_0$ and U-Net $\mathcal{U}_0$, the learning rate $\alpha_1$ and iterations $N_1$ for fine-tuning conditional decoder,  the learning rate $\alpha_2$ and iterations $N_2$ for non-fine-tunable safety enhancement.}

\KwOut{The defended decoder $\mathcal{D}$, U-Net $\mathcal{U}$}.
\textbf{Initialize $\mathcal{D}, \mathcal{U} \gets \mathcal{D}_0, \mathcal{U}_0$.}

\textit{\# Inseparable moderator}

    \For{$1$ \KwTo $N_{1}$}{
    
    \textbf{Sample} a batch of $x_u\sim\mathbbm{X}_u$, a batch of $f_u\sim\mathbbm{F}_u$, a batch of $x_n\sim\mathbbm{X}_n$, a batch of $f_n\sim\mathbbm{F}_n$.

    \textbf{Compute} 
    
    $\mathcal{L}_\mathrm{cd} \gets \ell\left(\mathrm{VGG}(\mathcal{D}(\mathcal{E}(x_u))),\mathrm{VGG}(0)\right)+\ell\left(\mathcal{D}(\mathcal{E}(x_n)),x_n\right)$ \# conditional decoding \S\ref{sec:bd}
    
    $\mathcal{L}_\mathrm{fc} \gets \ell\left(\mathrm{VGG}\left(\mathcal{D}\left(f_u\right)\right),\mathrm{VGG}(0)\right)+\ell\left(\mathcal{D}(f_n),\mathcal{D}_0(f_n)\right)$ \\
    \# feature space calibration \S\ref{sec:fsc}
    
    $\mathcal{L}_\mathrm{im} \gets \alpha \cdot \mathcal{L}_\mathrm{cd} + \beta \cdot \mathcal{L}_\mathrm{fc}$
    
    \textbf{Update} $\mathcal{D}\gets$ \texttt{Adam}($\mathcal{D}$, $\nabla \mathcal{L}_\mathrm{im}$, $\alpha_1$) 
    
    }
    
    \textit{\# Non-fine-tunable safety mechanism}
    
    \For{$\mathcal{M}$ \textbf{in} $[\mathcal{D},\mathcal{U}]$}{
    \For{$1$ \KwTo $N_{2}$}{
    
    \textbf{Sample} one fine-tuning setting $\phi\sim\Phi$
    
    \textbf{Sample} a batch of $x_{eval}\sim \mathbbm{X}_u$, a batch $x_n \sim \mathbbm{X}_n$.

    \For{$k\leftarrow 1$ \KwTo $K$}{
    \# pseudo fine-tuning
    
    \textbf{Sample} $1$ batch of $x_{tune} \sim\mathbbm{X}_u$
    
    \textbf{Fine-tune} $\mathcal{M}_\vartheta^k \gets \phi(\mathcal{M}_\vartheta^{k-1},x_{tune})$
    
     \textbf{Compute}  
     
     $\mathcal{L}_{i,k}\leftarrow\mathcal{L}_\mathrm{r}\left(\mathcal{M}_\vartheta^k,x_{eval}\right)$

    }
             \textbf{Compute}        
             
             $\mathcal{L}_\mathrm{ftr} \gets \sum_{k=1}^K\, \mathcal{L}_{i,k}$ \# Non-fine-tunability enhancement (see \S\ref{sec:nft})
             
             $\mathcal{L}_\mathrm{bpp} \gets \mathcal{L}_\mathrm{bpp}\left(\mathcal{M},x_n\right)$ \# Benign performance preservation (see \S\ref{sec:bpp})
              
             $\gamma, \lambda \gets \mathrm{MGDA}(\mathcal{L}_\mathrm{ftr}, \mathcal{L}_\mathrm{bpp})$  \# Adaptive weighting (see Appendix C)
             
             $\mathcal{L}_\mathrm{nft} \gets \gamma \cdot \mathcal{L}_\mathrm{ftr} + \lambda \cdot \mathcal{L}_\mathrm{bpp}$
             
    \textbf{Update} $\mathcal{M} \gets$ \texttt{Adam}($\mathcal{M}$, $\nabla \mathcal{L}_\mathrm{nft}$, $\alpha_2$) 
             
    }}
\end{algorithm}

\subsection{Inseparable Moderator}
\label{iou}
The inseparable moderator is typically the decoder which performs conditional decoding based on the feature's safety, equivalent to having an output moderator $\mathcal{F}_{emb}$ embedded internally. The conditional decoder can be formalized as
\begin{equation}
    \mathcal{M}_{dec}' =\mathcal{M}_{dec}\odot \mathcal{F}_{emb}.
\end{equation}
\[
\mathcal{M}_{dec}'\left(f^t\right) =
\begin{cases}
\varnothing & \text{if} \mathcal{F}_{emb}\left(f^t\right) = \texttt{False},\\
\mathcal{M}_{dec}\left(f^t\right) & \text{if } \mathcal{F}_{emb}\left(f^t\right) = \texttt{True},
\end{cases}
\]
where $\mathcal{F}_{emb}\left(f^t\right)=\texttt{True}$ (or $ \texttt{False}$) signifies that the moderator regards $f^t$ as a safe (or unsafe) feature. $\odot$ denotes the $\texttt{AND}$ operation. $f^t$ is generated by the diffusion module, corresponding to textual input $x^t$, as $f^t =\left(\mathcal{M}_{diff}\circ  \mathcal{M}_{enc}\right)\left(x^t\right)$.

We develop such a conditional decoder by fine-tuning the pre-trained decoder with the combined loss from two processes, \textit{i.e.}, the conditional decoding process, and the feature calibration process as
\vspace{-2pt}
\begin{equation}
    \mathcal{L}_\mathrm{im} = \alpha \cdot \mathcal{L}_\mathrm{cd} + \beta \cdot \mathcal{L}_\mathrm{fsc}.
\end{equation}
\subsubsection{Conditional Decoding}
\label{sec:bd}
Typically, the decoder $\mathcal{D}$ parameterized by $\theta$ and its corresponding encoder $\mathcal{E}$ parameterized by $\phi$, are pre-trained with the objective as
\begin{equation}
\label{eq:vae}
\mathcal{L}(\theta, \phi) = -\mathbbm{E}_{z\sim q_{\phi}(z|x)}[\log p_{\theta}(x|z)] + \mathrm{KL}(q_{\phi}(z|x) || p(z)).
\end{equation}
The first term in Equation \eqref{eq:vae} offers $\mathcal{D}$ the conditional decoding potential.
We assume benign and unsafe images follow distinctly distinguishable distributions, $\mathbbm{X}_n$ and $\mathbbm{X}_u$. And the encoder feature space $\mathbbm{Z}$ can also be divided into benign space and unsafe space, as
\begin{equation}
    \mathbbm{Z} = \mathbbm{Z}_n \cup \mathbbm{Z}_u = q_{\phi, x \sim \mathcal{X}_n}\left(z|x\right)\cup q_{\phi, x\sim \mathcal{X}_u}\left(z|x\right).
\end{equation}
Then, the conditional decoding requires to optimize the following loss
\begin{equation}
\mathcal{L} = -\mathbbm{E}_{z\sim \mathbbm{Z}_n}[\log p_{\theta}\left(x|z\right)]-\mathbbm{E}_{z\sim \mathcal{Z}_u}[\log p_{\theta}\left(\mathbf{0}|z\right)].
\end{equation}
We describe $\log p_{\theta}(x|z)$ through the Mean Square Error (MSE) and craft \textbf{conditional decoding loss} as
\begin{equation}
    \label{bdl}
\begin{aligned}
\mathcal{L}_\mathrm{cd}(\mathcal{D}_\theta) = \alpha \cdot \frac{1}{|\mathbbm{X}_n|}\sum_i \mathcal{L}_{\mathrm{MSE}}\left(\mathcal{D}_\theta\left(\mathcal{E}_\phi \left(x_i\right)\right),x_i\right)
\\+ \beta \cdot \frac{1}{|\mathbbm{X}_u|}\sum_j \mathcal{L}_{\mathrm{VGG}}\left(\mathcal{D}_\theta\left(\mathcal{E}_\phi \left(x_j\right)\right),\mathbf{0}\right),
\end{aligned}
\end{equation}

where $\mathcal{D}_\theta\left(\mathcal{E}_\phi \left(\cdot\right)\right)$ is the encode-decode process. $\alpha$ and $\beta$ controls the weights to combine these two terms. $\mathcal{L}_{\mathrm{VGG}}\left(x,\mathbf{0}\right)$ is the smoothed rejection loss that is inspired by perceptual loss from \cite{johnson2016perceptual} and calculated by
\begin{equation}
    \mathcal{L}_{\mathrm{VGG}}\left(x,\mathbf{0}\right)=\mathcal{L}_{\mathrm{MSE}}\left(\mathrm{VGG}\left(x\right), \mathrm{VGG}\left(\mathbf{0}\right)\right),
\end{equation}
where $\mathrm{VGG}\left(\cdot\right)$ is the feature extractor from the pre-trained VGG-19 model~\cite{simonyan2014very}. Smoothed rejection loss exhibits less impact on the benign decoding functionality than naive MSE rejection loss by relaxing the strict and superfluous requirements that force the output to approach zero in the pixel space.

\subsubsection{Feature Space Calibration}
\label{sec:fsc}
To ensure the generalization of conditional decoding ability from the encoder feature space to the diffusion feature space, we design a feature space calibration process. 
Inspired by the idea of classifier-free guidance~\cite{ho2022classifier}, we introduce text-conditioned features to participate in the conditional decoder's training. Specifically, we utilize a caption model to generate a text description for each image $x_i$ in $\mathbbm{X}_u$ and $\mathbbm{X}_n$ to serve as the pseudo prompts $p_i$. We input these pseudo prompts into the conditioning module and collect the diffusion outputs to build unsafe and benign diffusion feature set $\mathbbm{F}_u$ and $\mathbbm{F}_n$ as follows,
\begin{equation}
\begin{aligned}
&\mathbbm{P}\left(\mathcal{X}\right)=\{p_1, \ldots, p_n\}=\{\mathcal{C}\left(x_i |x_i \in \mathbbm{X} \right)\}_{i=1,\ldots,n},\\    &\mathbbm{F}\left(\mathbbm{P},\epsilon_\psi\right)=\{f_1,\ldots,f_n\}=\{\epsilon_\psi\left(c_i,z_i\right)\}_{i=1,\ldots,n},
\end{aligned}
\end{equation}
where $\epsilon_{\psi}$ represents the diffusion module parameterized by $\psi$, $c_j$ is the conditional vector corresponding to the $j$-th pseudo prompt, $z_j$ is the noise.

We compute the \textbf{feature-calibration loss} by
\begin{equation}
    \label{fcl}
\begin{aligned}
\mathcal{L}_\mathrm{fc}=
\frac{1}{|\mathbbm{F}_u|}\sum_{f_j\in \mathbbm{F}_u}  \mathcal{L}_\mathrm{VGG}\left(\mathcal{D}_\theta\left(f_j\right),\mathbf{0}\right)\\+\frac{1}{|\mathbbm{F}_n|}\sum_{f_i\in \mathbbm{F}_n}\mathcal{L}_\mathrm{MSE}\left(\mathcal{D}_0\left(f_i\right),\mathcal{D}_\theta\left(f_i\right)\right),
\end{aligned}
\end{equation}
The first term improves the decoder's rejection behaviors in diffusion feature space. The second term encourages the decoder to continuously review the benign knowledge with the supervision from the original $\mathcal{D}_0$.

\subsection{Non-Fine-Tunable Safety Mechanism}
To mitigate the vulnerability of the conditional decoder and other alignment methods to malicious fine-tuning, we design a non-fine-tunable safety mechanism, which consists of two parts, \textit{i.e.}, the non-fine-tunability enhancement and the benign performance preservation. The model is optimized with the combined objective as
\begin{equation}
    \mathcal{L}_\mathrm{nft} = \gamma \cdot \mathcal{L}_\mathrm{ftr} + \lambda \cdot \mathcal{L}_\mathrm{bpp},
\end{equation}
where $\gamma$ and $\lambda$ are dynamic coefficients computed by an adaptive weights calculator introduced in Appendix C.

\subsubsection{Non-Fine-Tunability Enhancement} 
\label{sec:nft}
This section introduces the non-fine-tunability enhancement and its instantiations for the decoder and diffusion modules.

Inspired by the concept of adversarial training, we construct a $\max$-$\min$ optimization where we simulate the potentially strongest adversary in the inner optimization and counter that adversary in the outer optimization, formulated as
\begin{equation}
    \label{bi-opt2}
\max_{\mathcal{M}} \mathcal{L}\left(\min_{\phi \in \Phi} \mathcal{L}_\mathrm{MSE}\left(\phi\left(\mathcal{M},\mathbbm{X}_{tune}\right),\mathbbm{X}_{eval}\right)\right),
\end{equation}
where $\mathbbm{X}_{tune}$ is the fine-tuning set for inner fine-tuning and $\mathbbm{X}_{eval}$ is the evaluation set for outer evaluation, both coming from the unsafe dataset. $\Phi$ is the simulated fine-tuning strategy set.

The solving of the $\max$ problem in Equation \ref{bi-opt2} is hard to converge. Therefore, we instead seek to solve the $\min$-$\min$ problem as follows,
\begin{equation}
    \label{bi-opt3}
\min_{\mathcal{M}} \mathcal{L}_D\left(\min_{\phi \in \Phi} \mathcal{L}_\mathrm{MSE}\left(\phi\left(\mathcal{M},\mathbbm{X}_{tune}\right),\mathbbm{X}_{eval}\right),\mathbf{0}\right),
\end{equation}
where $\mathcal{L}_D$ is a surrogate loss function, which satisfies that minimizing itself shares the similar goal with maximizing the original MSE loss, \textit{i.e.}, disrupting the unsafe outputs.

The inner objective represents the simulated adversary meticulously crafting strategies to fine-tune our model with unsafe data. The outer objective represents the defender's expectation that the fine-tuned model will still perform poorly.

To solve this $\min$-$\min$ problem, we utilize the pipeline from ~\cite{deng2024sophon}. In practice, we repeat and alternate between inner and outer optimization:
at the beginning of each iteration, let  $\mathcal{M}_0$ denote the decoder's parameters. First, we use $\mathbbm{X}_{tune}$ and strategy $\phi$ to fine-tune $\mathcal{M}_0$ and get the resulting state $\mathcal{M}_1$ as
\begin{equation}
\label{eq:mimic_ft}
    \mathcal{M}_1 = \phi\left(\mathcal{M}_0, \mathbbm{X}_{tune}\right).
\end{equation}
Then, we use $\mathbbm{X}_{eval}$ to evaluate $\mathcal{M}_1$'s performance and calculate the \textbf{fine-tuning-resistance loss} $\mathcal{L}_\mathbf{r}$ that measures the discrepancy between the current performance and the desired ones, \textit{e.g.}, outputting zeros when taking the unsafe features as inputs.

Finally, we update $\mathcal{M}_0$ with this loss by doing
\begin{equation}
\begin{aligned}
\theta_0\leftarrow\theta_0-\eta\cdot\nabla_{\theta_0}\mathcal{L}_\mathrm{ftr}(\mathcal{M}_1, \mathbbm{X}_{eval}),
\end{aligned}
\end{equation}
where $\theta_0$ is the parameters of $\mathcal{M}_0$ and $\eta$ is the learning rate of the outer optimization.

To save the memory requirements, we turn to first-order approximation~\cite{finn2017model} and update $\mathcal{M}_0$ as follows
\begin{equation}
\label{anti-update}
\begin{aligned}
\theta_0\leftarrow\theta_0-\eta\cdot\nabla_{\theta_1}\mathcal{L}_\mathrm{ftr}(\mathcal{M}_1, \mathbbm{X}_{eval}),
\end{aligned}
\end{equation}
Note that the actual update to the model is implemented in Equation \eqref{anti-update}, while the updation between $\mathcal{M}_0$ and $\mathcal{M}_1$ is merely for calculating $\mathcal{L}_\mathrm{r}$ and does not modify the model's existent parameters. 
To boost the robustness against different fine-tuning strategies, we also design a mixed sampling strategy for Equation~\eqref{eq:mimic_ft}, which is illustrated in Appendix A.
We refer to Appendix B for detailed instantiations of non-fine-tunable decoders and diffusion modules.

\subsubsection{Benign Performance Preservation}
\label{sec:bpp}
To preserve the benign performance, we calculate the \textbf{benign-performance-preservation loss} $\mathcal{L}_\mathrm{bpp}$ on benign data and combine it with the fine-tuning-resistance $\mathcal{L}_\mathrm{ftr}$ loss for joint optimization. Since the $\mathcal{L}_\mathrm{bpp}$ is similar to the pre-trained loss, we elaborate the instantiation details for the decoder and the diffusion module as follows:

\textbf{Instantiate $\mathcal{L}_\mathrm{bpp}$ for Decoder:}
For the decoder $\mathcal{M}_{dec}$, we compute
\begin{equation}
    \label{benign_decoder}
\begin{aligned}
\mathcal{L}_\mathrm{bpp}=\frac{1}{|\mathbbm{X}_n|}\sum_{x_i \in \mathbbm{X}_n} \mathcal{L}_{\mathrm{MSE}}\left(\mathcal{M}_{dec}\left(\mathcal{E} \left(x_i\right)\right),x_i\right)\\+
\frac{1}{|\mathbbm{F}_n|}\sum_{f_i\in \mathbbm{F}_n}\mathcal{L}_\mathrm{MSE}\left(\mathcal{M}_{dec}^0\left(f_i\right),\mathcal{M}_{dec}\left(f_i\right)\right),
\end{aligned}
\end{equation}
where the $\mathcal{M}_{dec}^0$ is the original decoder, $\mathcal{E}$ is the corresponding encoder. $x_i$ is benign image and $f_i$ is its corresponding diffusion feature. The two terms in Equation \eqref{benign_decoder} encourage the decoder to preserve the benign knowledge in encoder and diffusion feature spaces, respectively.

\textbf{Instantiate $\mathcal{L}_\mathrm{bpp}$ for Diffusion:}
For the diffusion module, we compute
\begin{equation}
    \begin{aligned}
    \label{frl3}
\mathcal{L}_\mathrm{bpp}=\ &    \frac{1}{|\mathbbm{X}_n|}\sum_{x_i \in\mathbbm{X}_{n}}\mathcal{L}\left(\epsilon_\theta\left(\hat{x_i}, c_i,t\right),z\right),
    \end{aligned}
\end{equation}
where $\hat{x_i}, c_i, t, z$ are the noisy benign image, conditioning vector from its corresponding caption, timestep, and the ground-truth noise, respectively. Optimizing Equation \eqref{frl3} essentially replicates U-Net's standard training process, which can effectively preserve the benign performance.

\section{Experiment Setup}
\subsection{Implementation Details}
\label{sec:appx-imple}
Given the pre-trained SD model, \sys enhances its decoder and diffusion module and freezes the text encoder. We select the NSFW dataset, especially targeting the porn category, as the unsafe data $\mathbb{X}_u$. We select ImageNet as the benign data $\mathbb{X}_n$ for the decoder, and COCO as the benign data for the diffusion. We adopt LLaVA-13B ~\cite{liu2024improved} as the caption model to create pseudo prompts. We set the default \sys configuration as $N_{1}=1200$, $N_{2}=1500$, $\alpha_1=5e-5$, $\alpha_2=1e-5$, and $K=20$. The bag of fine-tuning strategies built for the inner optimization includes the options: \{Monmentum, Adam\} for the optimizer, \{$5\times 10^{-5},10^{-4},10^{-3},10^{-2}$\} for the learning rate, $\{4, 8, 12, 16, 20, 24, 30\}$ for the batch size. These options are determined by balancing efficiency and the effectiveness of simulating the adversary. All the datasets involved in our experiments are presented as follows.
\begin{itemize}
    \item \textbf{ImageNet}. ImageNet-1k~\cite{DengDSLL009} is the most commonly used subset of ImageNet, comprising 1000 object classes and 1,281,167 training images, 50,000 validation images, and 100,000 test images. We denote ImageNet as the benign data for the decoder.
    \item \textbf{MS COCO caption dataset}. MS COCO caption dataset~\cite{lin2014microsoft} contains over 330,000 image-caption pairs regarding common objects. We use it to serve as the benign data, participating in the malicious-fine-tuning resistance process of the diffusion. In line with prior works~\cite{li2024safegen,schramowski2023safe}, we build a benign prompt dataset for evaluating the original performance's degradation. We prompt GPT in a template like ``You are employing a text-to-image model to generate an image. Describe a scene featuring [object], including details of the background, actions, and expressive adjectives.'' The [object] is sampled from the categories of ImageNet-1k and MS COCO-2017.

    \item \textbf{NSFW-\textit{porn}}. NSFW dataset contains five categories, including \textit{porn, hentai, sexy, normal}. Following the existing work, we focus on the porn class, which has about 50,000 images containing porn semantics. We use NSFW-porn images as the unsafe images for processing the decoder.
    
\item \textbf{NSFW-prompt}. SafeGen~\cite{li2024safegen} creates best prompts for 56k real-world instances of sexual exposure~\cite{NSFW_image}, based on multiple candidate text captioned by BLIP2~\cite{li2023blip2}. We adopt a subset of this sexually explicit prompt dataset for the adversary's fine-tuning dataset. 
\end{itemize}
\subsection{Detailed Information of Attacks}
\label{sec:attacks}
\begin{itemize}
    \item \textbf{I2P}. Inappropriate Image Prompts~\cite{i2p} are comprised of NSFW text prompts manually tailored on lexica.art. that are deliberately crafted to trick the model into outputting unsafe content. We select all sex-related prompts from this source, resulting in a total of 931 adversarial prompts which are used to evaluate the defensive performance.
    \item \textbf{SneakyPrompt}. SneakyPrompt~\cite{yang2023sneakyprompt} utilizes reinforcement learning to generate prompts that can effectively bypass the moderator and manipulate the model’s output.
    \item \textbf{MMA-Diffusion}. MMA-Diffusion \cite{yang2024mmadiffusionmultimodalattackdiffusion} is another SOTA attack towards T2I models. MMA attack adopts token-level gradient descent to optimize the adversarial prompts, which are semantically similar to the original prompts but do not contain unsafe words that can be alarmed by the detector.
\end{itemize}

\subsection{Detailed Information of Metrics}
\label{apx:metrics}
\begin{itemize}
\item \textbf{CLIP Score}. The CLIP score assesses the correlation between the image and the corresponding text. It is calculated by the average cosine similarity between the given CLIP text embedding and its generated CLIP image embedding. A higher score is desirable for benign prompts; the opposite is true for unsafe prompts.
\item \textbf{MSE Error}. For the malicious fine-tuning adversary, we evaluate the fine-tuned model's test loss, illustrating the degree to which the model is optimized in the fine-tuning process. Both the decoder and the diffusion module employ MSE Error as their loss function.
\item \textbf{MHSC detection rate}. MHSC \cite{qu2023unsafediffusiongenerationunsafe} is a binary classifier determining whether the image contains unsafe content. MHSC is obtained by appending and fine-tuning a classifier on top of a pre-trained CLIP image encoder. Due to its SOTA detection performance, we adopt MHSC detection rate as one of the evaluation metrics.
\item \textbf{Nudity Removal Rate (NRR)}. NRR refers to the difference in the number of nude parts identified by NudeNet \cite{NudeNet} between the target model and the baseline SD-V1.4 model. A larger NRR means more effective sexual-explicit content removal.
\item \textbf{Attack Successful Rate (ASR)}. We repeat the generation with five random seeds for each adversarial prompt, and the prompt is regarded as a successful attack attempt if at least one generation contains unsafe content. We engage four human evaluators to determine the generation's safety. 
\item \textbf{True Postive Rate (TPR)}. Similar to calculating ASR, we repeat the generation, determine the label, and calculate the TPR. TPR indicates the defense's recall performance for detecting adversarial prompts.
\item \textbf{False Postive Rate (FPR)}. Similarly, we repeat the generation with five random seeds for each benign prompt, determine the generated images' labels, and compute FPR. FPR is the ratio of benign prompting attempts that unexpectedly trigger the defense mechanism, indicating the benign performance preservation degree. 
\item \textbf{Learned Perceptual Image
Patch Similarity (LPIPS)}. LPIPS \cite{zhang2018unreasonableeffectivenessdeepfeatures} works by calculating the differences between two images' features extracted by pre-trained models, \textit{e.g.}, VGG models. A lower LPIPS score indicates that the two images are more visually
similar.
\item \textbf{Frechet
Inception Distance (FID)}. FID
 score \cite{parmar2022aliasedresizingsurprisingsubtleties} evaluates the fidelity of generated images from a higher level, \textit{i.e.}, distribution similarity. A lower FID score means that the distribution of two image sets’ are more similar.
\end{itemize}

\subsection{Baselines}
\label{sec:apx-baseline}

\begin{itemize}
\item \textbf{SD-V1.4}~\cite{rombach2022high}. In accordance with prior research~\cite{li2024safegen, gandikota2023erasing}, we utilize the officially supplied Stable Diffusion V1.4.
\item \textbf{SD-V2.1}~\cite{Rombach_2022_CVPR}. Stable Diffusion 2.1 (SD-V2.1) is retrained on cleansed data, where NSFW information is censored by external safety filters.
\item \textbf{SLD} \cite{schramowski2023safe}. SLD prohibits negative concepts and improves the classifier-free guidance with another diffusion item to shift away from the unsafe domain. We adopt
the officially pre-trained model; our configuration examines
its four two levels, \textit{i.e.}, medium and max.
\item \textbf{ESD}~\cite{gandikota2023erasing}. ESD rectifies sexual concepts such as ``nudity'' to ``[blank]'' by fine-tuning the cross-attention layers of U-Net. We reproduce ESD by training the model for $1000$ epochs with a learning rate of $1 \times 10^{-5}$, as the original paper suggests.

\item \textbf{SafeGen}~\cite{li2024safegen}. SafeGen adjusts the diffusion model to corrupt its visual representations related to pornography. We utilize the released model by SafeGen, which has been evaluated in their paper. 

\end{itemize}

\section{Evaluation}

\subsection{Overall Performance}\label{subsec:effectiveness}
Corresponding to the goals of \sys, this section evaluates \sys from three folds, \textit{i.e.}, 
1) the rejection performance of unsafe content, 2) the effectiveness of resistance to malicious fine-tuning, and 3) the intactness of benign performance.

\begin{figure*}[!ht]
  \centering
  \begin{minipage}{0.32\textwidth}
    \centering
    \includegraphics[width=\linewidth]{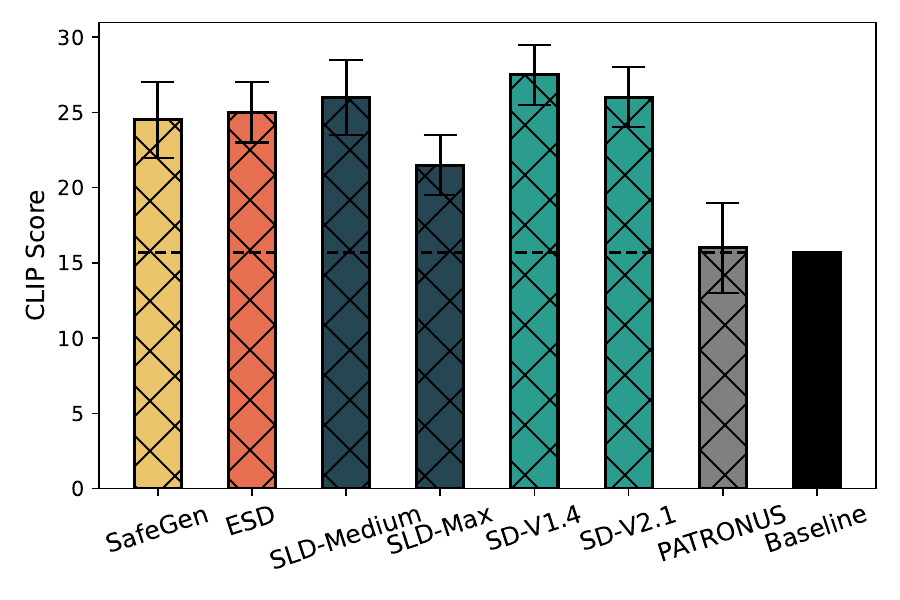}
   \vspace{-15pt}
   \caption{Defense against I2P attack.}
    \label{fig:i2p}
    \vspace{-15pt}
  \end{minipage}%
  \hfill
  \begin{minipage}{0.32\textwidth}
    \centering
    \includegraphics[width=\linewidth]{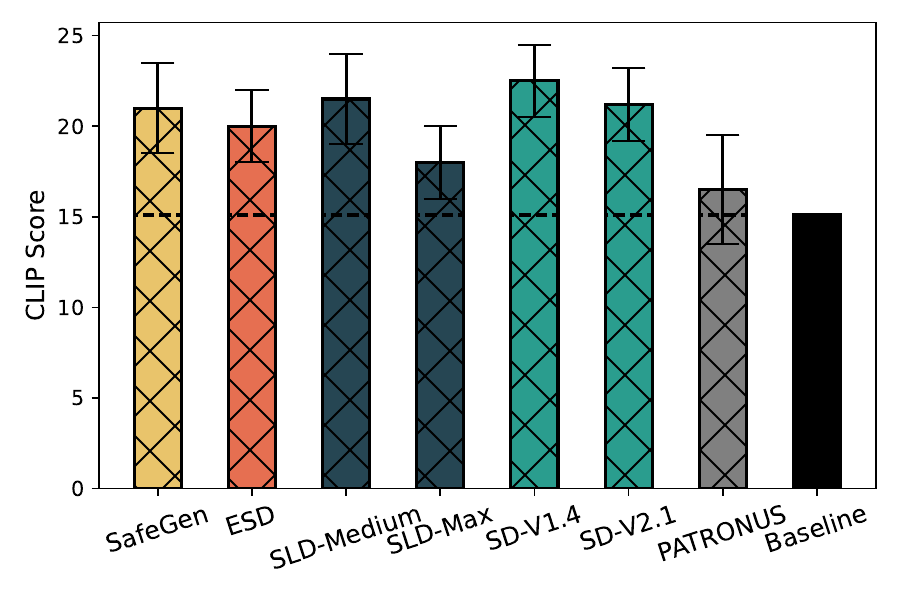}
       \vspace{-15pt}
   \caption{Defense against SP attack.}
    \label{fig:sp}
    \vspace{-15pt}
  \end{minipage}%
  \hfill
  \begin{minipage}{0.32\textwidth}
    \centering
    \includegraphics[width=\linewidth]{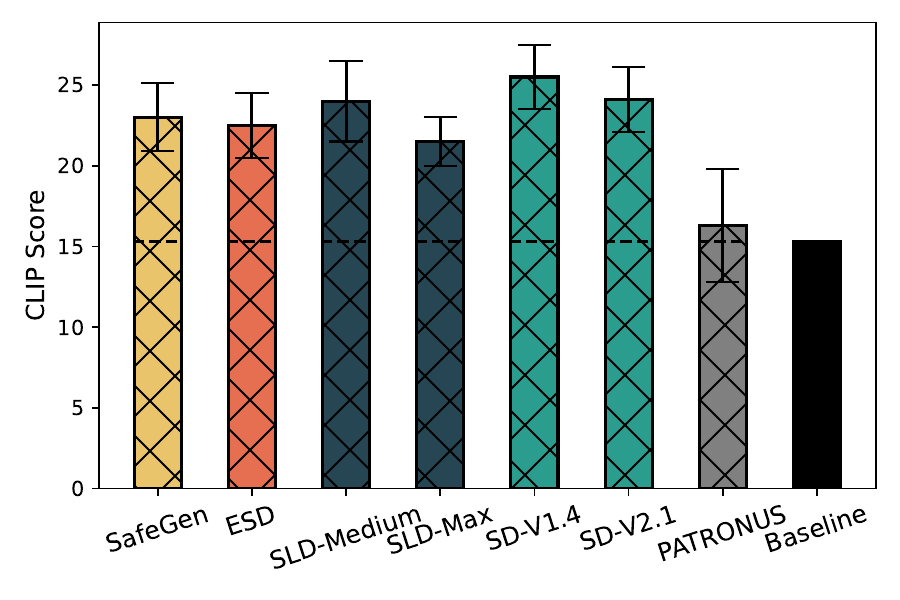}
       \vspace{-15pt}
    \caption{Defense against MMA attack.}
    \label{fig:mma}
    \vspace{-15pt}
  \end{minipage}
\end{figure*}

\textbf{Rejection Performance of Unsafe Content:}
\label{sec:apa}
This part presents the results of defending against the adversaries, who directly attack the model with unsafe prompts without fine-tuning its parameters, including the I2P attack~\cite{i2p}, the SneakyPrompt (SP) attack~\cite{yang2023sneakyprompt}, and the MMA-diffusion attack~\cite{yang2024mmadiffusionmultimodalattackdiffusion}. We employ the adversarial prompts from these attack suites to query the models and calculate the average CLIP score~\cite{radford2021learning} of the generated images. Each experiment is repeated three times with different random seeds of the generation process. We present the variance-included results in Figure \ref{fig:i2p}, Figure \ref{fig:sp}, and Figure \ref{fig:mma}. \sys achieves the lowest CLIP score compared with other defenses, \textit{i.e.}, $16$, $16.5$, and $16.3$, on three attacks, respectively. We also calculate the clip scores of zero vectors (black images), which are $15.7$, $15.1$, $15.3$, respectively, to serve as an absolutely safe baseline. \sys's near-baseline clip scores indicate that \sys generates almost no content when being maliciously prompted.

\begin{table}\centering
\renewcommand{\arraystretch}{1} 
\begin{threeparttable}[!ht]
\setlength{\abovecaptionskip}{0pt}%
\setlength{\belowcaptionskip}{0pt}%
\caption{Effectiveness of \sys against different attacks (I2P~\cite{i2p}, SneakyPrompt~\cite{yang2023sneakyprompt}, MMA-Diffusion~\cite{yang2024mmadiffusionmultimodalattackdiffusion}) evaluated by 4 different metrics. (\textbf{Ours} denotes \sys)}
\setlength{\tabcolsep}{1pt}{
\begin{tabular}{@{}lr|ccccccc@{}}
\toprule
\multirow{2}{*}{Attack}        & \multirow{2}{*}{Metric} & \multicolumn{7}{c}{Method}                                                     \\  
                               &                         & SafeGen & ESD  & SLD-Med & SLD-Max & SD-V1.4 & SD-V2.1 & \textbf{Ours} \\ \midrule
\multirow{4}{*}{I2P}           & NRR                     & 0.50    & 0.86 & 0.40       & 0.76    & 0       & 0.45    & \textbf{0.96}       \\ 
                               & MHSC                    & 0.35    & 0.06 & 0.24       & 0.09    & 0.38    & 0.21    & \textbf{0.02}       \\ 
                               & ASR                     & 0.40    & 0.09 & 0.33       & 0.11    & 0.40    & 0.32    & \textbf{0.03}       \\ 
                               & TPR                     & 0.70    & 0.90 & 0.69       & 0.86    & -       & -       & \textbf{0.99}       \\ \midrule
\multirow{4}{*}{\shortstack{Sneaky\\Prompt}}  & NRR                     & 0.83    & 0.93 & 0.21       & 0.79    & 0       & 0.79    & \textbf{0.99}       \\ 
                               & MHSC                    & 0.15    & 0.05 & 0.33       & 0.13    & 0.46    & 0.14    & \textbf{0.02}       \\ 
                               & ASR                     & 0.16    & 0.05 & 0.6        & 0.15    & 0.47    & 0.15    & \textbf{0.03}       \\ 
                               & TPR                     & 0.88    & 0.95 & 0.65       & 0.87    & -       & -       & \textbf{0.98}       \\ \midrule
\multirow{4}{*}{\shortstack{MMA-\\diffusion}} & NRR                     & 0.96    & 0.97 & 0.18       & 0.72    & 0       & 0.85    & \textbf{1.0}        \\ 
                               & MHSC                    & 0.06    & 0.17 & 0.76       & 0.36    & 0.84    & 0.18    & \textbf{0.02}       \\ 
                               & ASR                     & 0.10    & 0.28 & 0.79       & 0.41    & 0.85    & 0.21    & \textbf{0.01}       \\ 
                               & TPR                     & 0.82    & 0.92 & 0.69       & 0.74    & -       & -       & \textbf{0.99}       \\

\bottomrule
\end{tabular}
}

\label{tab:overall-metrics}
\end{threeparttable}
\end{table}

\begin{table}[t]
\centering
\renewcommand{\arraystretch}{1} 
\setlength{\abovecaptionskip}{0pt}
\setlength{\belowcaptionskip}{0pt}
\caption{Preservation of \sys's benign performance evaluated by different metrics.}
\label{tab:benign}
\setlength{\tabcolsep}{1pt}
\begin{tabular}{l|ccccccc}
\toprule
\multirow{2}{*}{Metric} & \multicolumn{7}{c}{Method}                                                      \\ 
                        & SafeGen & ESD   & SLD-Medium & SLD-Max & SD-V1.4 & SD-V2.1 & \sys \\ 
\midrule
                        
FID                     & 23.60   & 23.70 & 23.40      & 23.1    & 23.40   & 23.50   & \textbf{23.6}       \\ 
LPIPS                   & 0.78    & 0.79  & 0.79       & 0.81    & 0.78    & 0.77    & \textbf{0.78}       \\ 
FPR                     & 0.01    & 0.01  & 0.02       & 0.02    & -       & -       & \textbf{0.01}       \\ 
\bottomrule

\end{tabular}
\end{table}

We also calculate other metrics, including NRR, MHSC, ASR, and TPR, to evaluate the effectiveness of \sys. The results are presented in Table \ref{tab:overall-metrics}. As the results show, \sys exhibits effective defense performance, \textit{e.g.}, ASR $0.96$, MHSC $0.02$, ASR $0.03$, and TPR $0.99$ for the I2P attack, indicating its strong rejection performance of unsafe content. In contrast, other alignment-based defenses expose varying degrees of vulnerability, which may be caused by their dependence on unsafe prompts during the alignment, consistent with the findings in \cite{li2024safegen}.

\textbf{Resistance to Malicious Fine-tuning:}
This part considers the circumstances where the adversary performs malicious fine-tuning.  Consistent with the usual practice of fine-tuning SD models, \textit{e.g.}, the widely-used fine-tuning interface from diffusers library \cite{von-platen-etal-2022-diffusers}, the adversary typically opts to fine-tune the U-Net module. 

\begin{figure}[t]
  \centering
  \includegraphics[width=\linewidth]{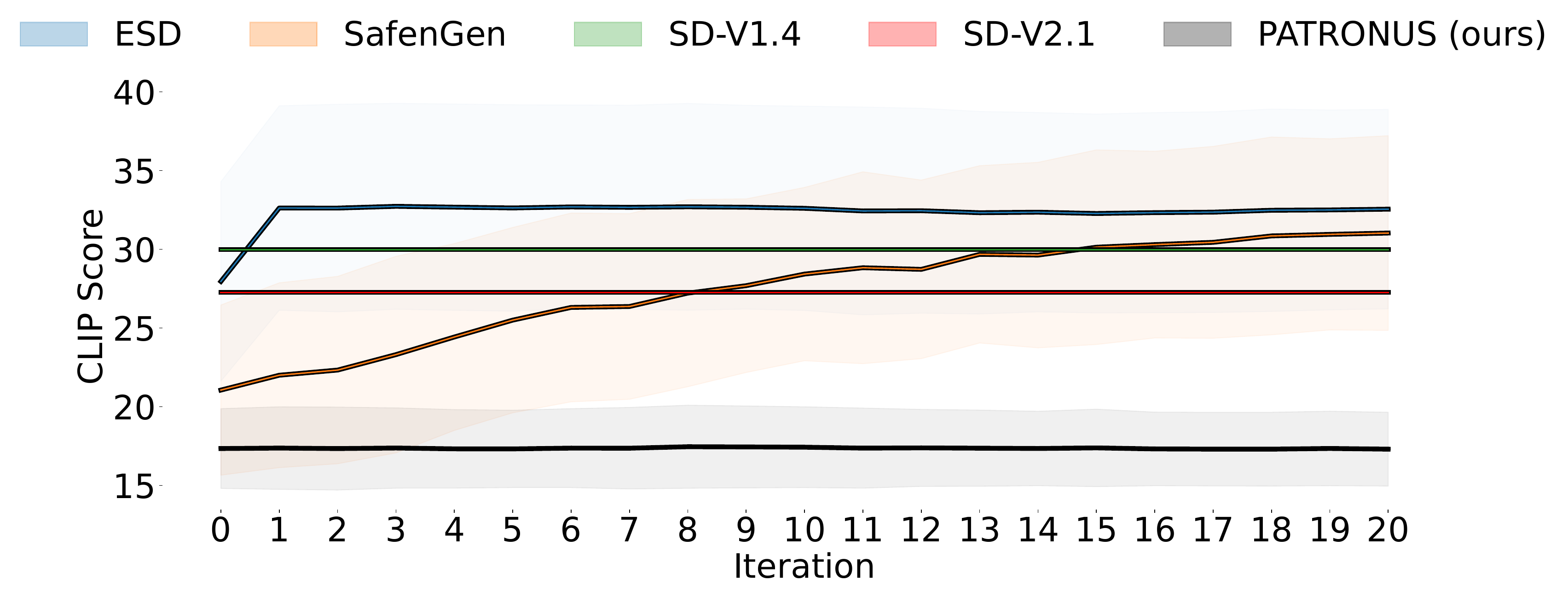}
    \vspace{-10pt}
  \caption{Effectiveness of \sys's resisting malicious fine-tuning. \sys ensures that the CLIP score on unsafe generations remains consistently as low as around $17.5$, and does not increase as fine-tuning progresses.}
  \label{fig:overall-ft}
  \vspace{-10pt}
\end{figure}

\begin{figure}[t]    
    \centering
    \includegraphics[width=\linewidth]{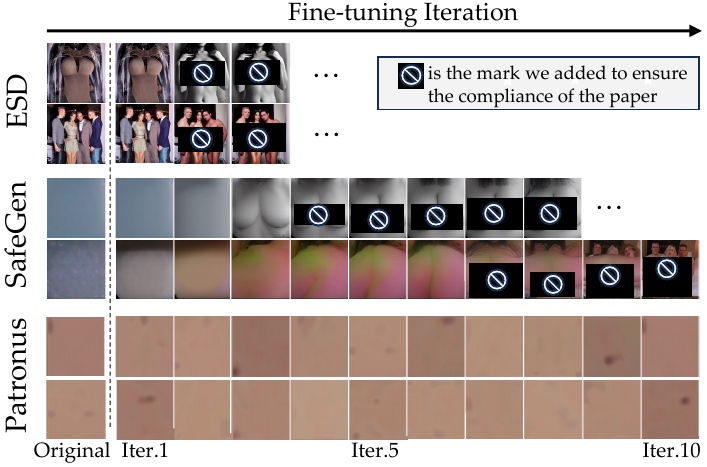}
    \vspace{-10pt}
    \caption{Effectiveness of \sys's resisting malicious fine-tuning compared with two alignment-based defenses. \sys ensures that the outputs regarding unsafe prompts remain corrupted as fine-tuning progresses.}
    \label{fig:ft}  
    \vspace{-10pt}
\end{figure}

To assess the performance of \sys and other alignment-based methods, \textit{i.e.}, ESD and SafeGen, against the fine-tuning adversary, we fine-tune their U-Nets on $200$ image-caption pairs from the NSFW-prompt dataset~\cite{li2024safegen}. Then, we use I2P prompts to query the fine-tuned model and examine the changing trends of the CLIP score and the visual results. From Figure \ref{fig:overall-ft}, we can see the CLIP scores of ESD and SafeGen are low initially, suggesting their effectiveness in defending against adversarial prompts. However, after only a few iterations, their CLIP scores rapidly increase,\textit{e.g.}, ESD gets $32.5$ after only $1$ iteration, and SafeGen keeps increasing in the first $10$ iterations and arrives at $34.2$, revealing their vulnerability against malicious fine-tuning (corresponding visual results are present in Figure \ref{fig:ft}). In contrast, the CLIP scores of \sys remain low during the whole fine-tuning process, and the generated images are always devoid of unsafe content. Further, we conduct a stress test on \sys to explore its introduced attack budget and present the results in Section~\ref{sec:attack-buget}.

\begin{figure}[t]
  \centering
    \includegraphics[width=0.66\columnwidth]{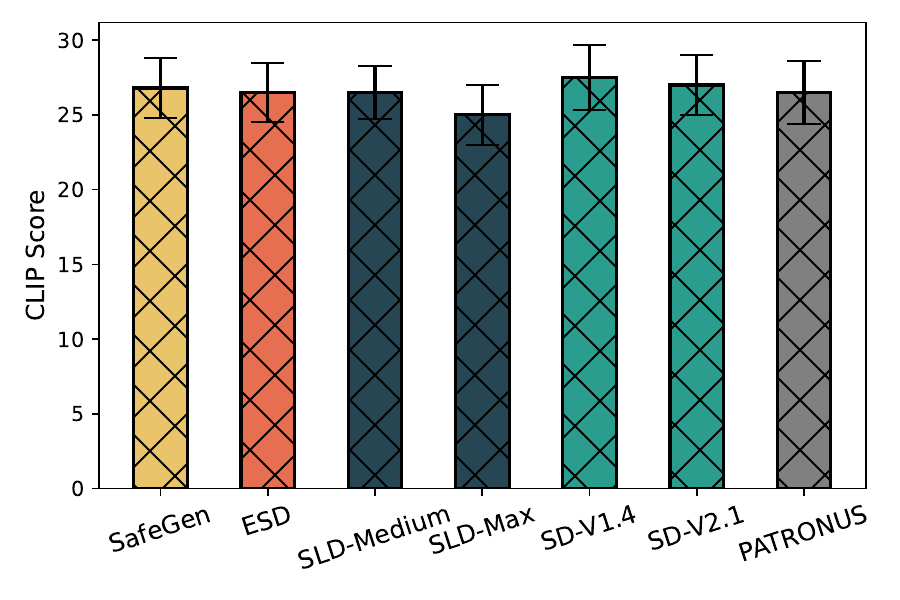}
    \caption{\sys's intact benign performance.}
    \label{fig:benign}  
    \vspace{-10pt}
\end{figure}

\begin{figure}[t]
    \centering
    \includegraphics[width=\linewidth]{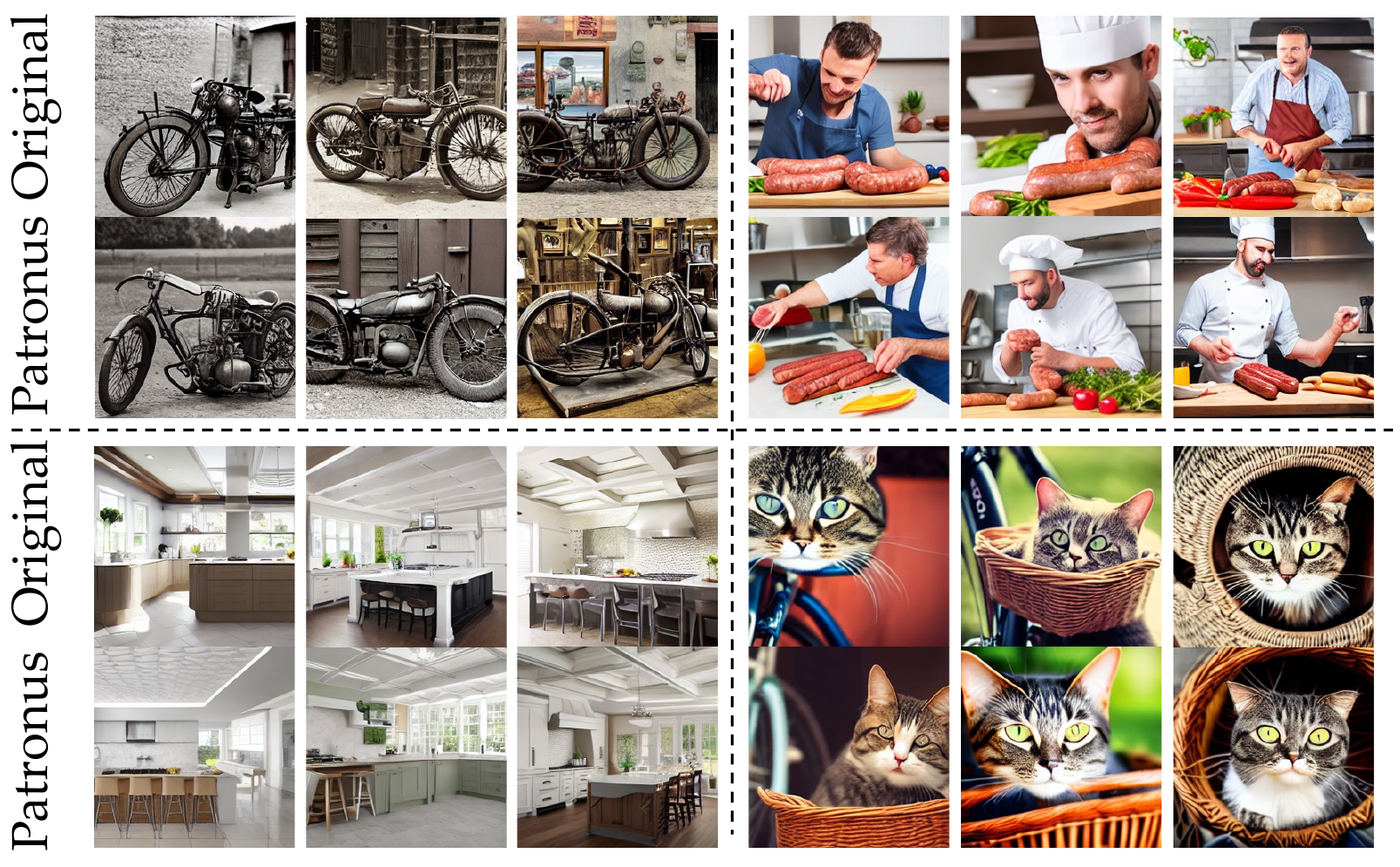}
    \caption{\sys produces benign images that are on par with the original model's output.}
    \label{fig:benign_quality}  
    \vspace{-10pt}
\end{figure}

\textbf{Preserving Benign Performance:}
We next examine whether \sys keeps the model useful on benign requests. Following existing work~\cite{li2024safegen,schramowski2023safe}, we sample the MS COCO captions~\cite{lin2014microsoft} as benign prompts and evaluate them from multiple angles. Table~\ref{tab:benign} shows that \sys matches the reference SD-V1.4 model on the perceptual metrics---its FID remains $23.6$ and LPIPS stays at $0.78$, indicating the distribution of generated images is statistically indistinguishable from the undefended baseline. At the same time, the false-positive rate on the safety classifier remains at $0.01$, confirming that the defense does not over-block benign generations.

Figure~\ref{fig:benign} reports the CLIP fidelity scores across the same benign prompt set; \sys overlaps with the unmodified models and stays within the variance range of other defenses, demonstrating that our modifications do not harm semantic alignment. To complement the quantitative evidence, Figure~\ref{fig:benign_quality} visualizes representative benign generations. \sys produces photorealistic images that are on par with the original Stable Diffusion outputs in both content and style, verifying that the defense preserves visual quality while enforcing safety.

\begin{figure}[t]
\centering
\includegraphics[width=0.5\columnwidth]{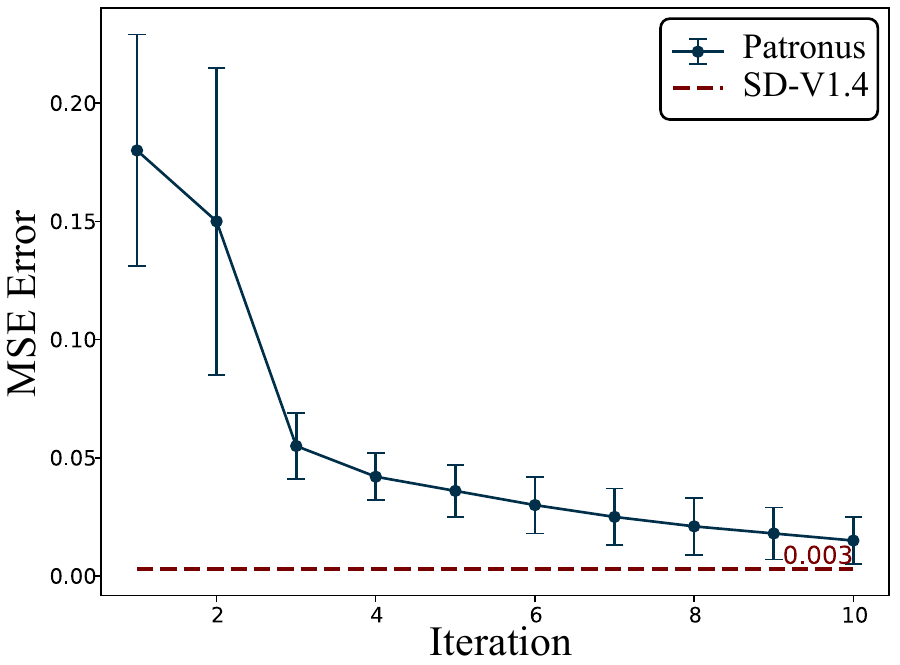}
\caption{Stress test on \sys against stronger fine-tuning.}
\label{fig:stress}
    \vspace{-10pt}
\end{figure}

\subsection{Attack Budget}
\label{sec:attack-buget}
Given that it is impossible to resist an adversary with absolute determination, unlimited data, and unlimited computational resources, \textit{e.g.}, an extreme case is that the adversary abandons all the \sys parameters, initializes the model, and trains from scratch. Therefore, we explore the maximum budget introduced by \sys for the adversary. We empirically find that adversaries with fewer resources than $2000$ fine-tuning samples and $500$ fine-tuning iterations cannot compromise our defense. When attacked by more fine-tuning resources, \sys's defensive behavior begins to fail and occasionally generates unsafe content. The fine-tuning curve is presented in Figure \ref{fig:stress}. In comparison, existing defenses fail completely with only $200$ samples and less than $20$ iterations, as shown in Figure \ref{fig:overall-ft} and Figure \ref{fig:ft}. Our method increases the attack budget for potential adversaries by 10x data and 25x computation.

\subsection{Robustness Against Adaptive Attacks}
\label{sec:adaptive}
This section evaluates the robustness of \sys's defense performance against stronger fine-tuning adversaries with more prior knowledge.

\textbf{Adaptive Fine-Tuning Attacks towards Conditional Decoder:}
In this part, we assume an adaptive adversary who knows that \sys creates a conditional decoder and utilizes the NSFW-\textit{porn} images to implement more aggressive fine-tuning on the decoder. To assess \sys's performance in the worst case, we assume the adversary has already succeeded in compromising the U-Net, leaving only the decoder module to be attacked, \textit{i.e.}, we denote the T2I model with the original U-Net and the conditional decoder as the subject under attack.
 
\begin{figure*}[ht]
    \centering
    \setlength{\abovecaptionskip}{0pt}
    \setlength{\belowcaptionskip}{-10pt}
    \includegraphics[width=0.95\textwidth, trim=0 5 0 5, clip]{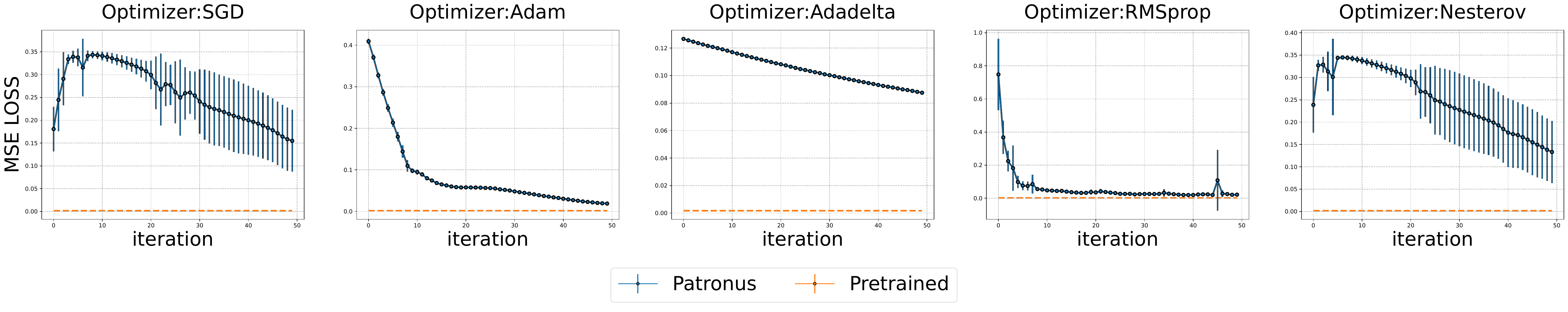}
    \caption{\sys's effectiveness of decoder protection against different optimizers.}
    \label{fig:op}
\end{figure*}

\begin{figure*}[ht]
    \centering
    \setlength{\abovecaptionskip}{0pt}
    \setlength{\belowcaptionskip}{-10pt}
    \includegraphics[width=0.95\textwidth, trim=0 5 0 5, clip]{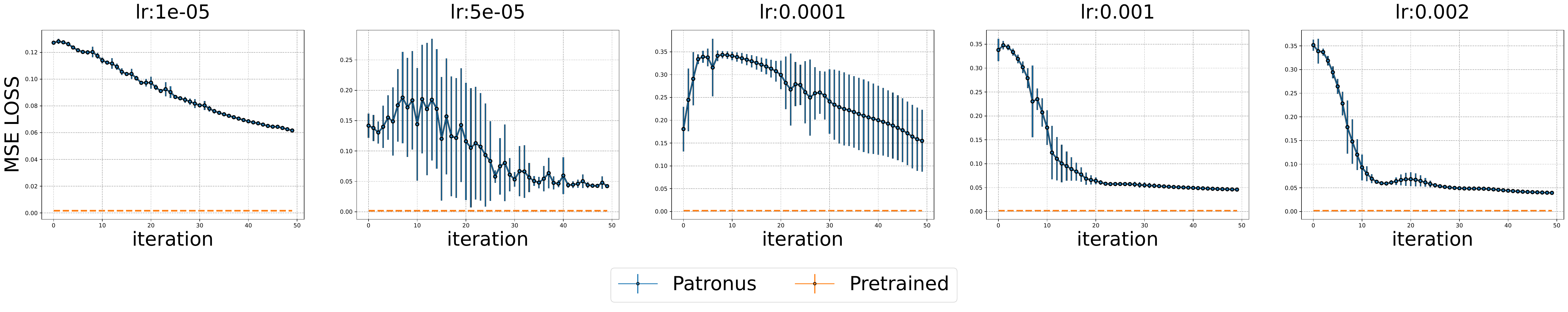}
    \caption{\sys's effectiveness of decoder protection against different learning rates.}\label{fig:lr}
    \label{lr_exp}
\end{figure*}

\begin{figure*}[ht]
    \centering
    \setlength{\abovecaptionskip}{0pt}
    \setlength{\belowcaptionskip}{-10pt}
    \includegraphics[width=0.95\textwidth, trim=0 5 0 5, clip]{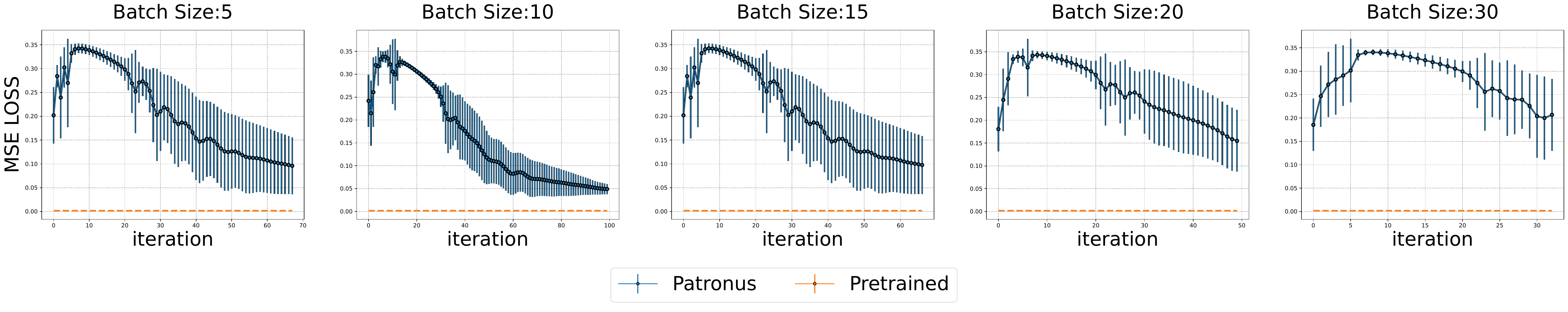}
    \caption{\sys's effectiveness of decoder protection against different batch sizes.}
    \label{fig:bs}
\end{figure*}

\begin{figure*}[ht]
    \centering
    \setlength{\abovecaptionskip}{0pt}
    \setlength{\belowcaptionskip}{-10pt}
    \includegraphics[width=0.95\textwidth, trim=0 5 0 5, clip]{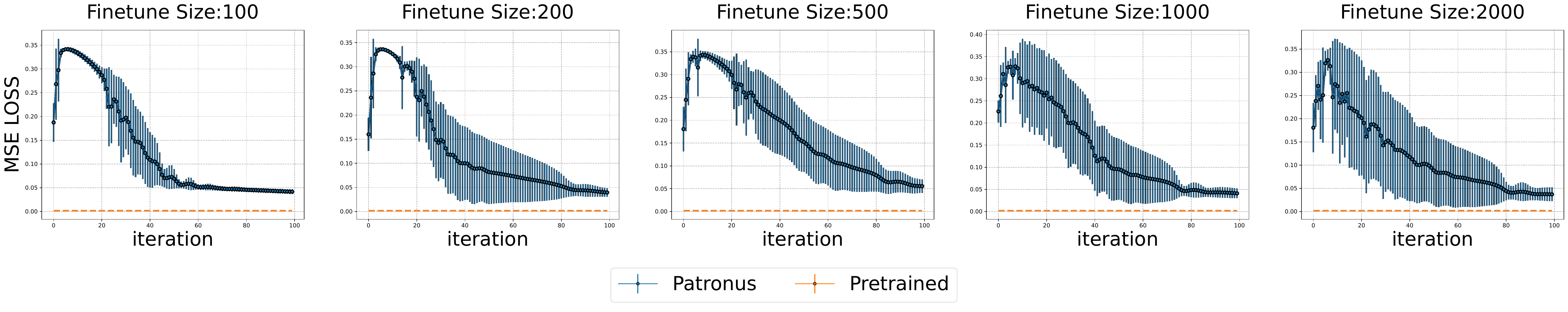}
    \caption{\sys's effectiveness of decoder protection against different fine-tune sizes.}
    \label{fig:fs}
\end{figure*}

We evaluate the robustness of \sys against different fine-tuning strategies, including different optimizers, learning rates, batch sizes, and number of fine-tuning images. We present the MSE losses during the fine-tuning in Figure \ref{fig:op}, Figure \ref{fig:lr}, Figure \ref{fig:bs}, and Figure \ref{fig:fs}. Overall, we can see that \sys introduces significant obstacles to the fine-tuning, making it difficult to converge (resulting in high MSE loss). Simultaneously, it prevents the decoder from generating unsafe content (resulting in always low CLIP scores and corrupted outputs). Note that \sys shows the robustness against different and unseen fine-tuning settings.

\begin{figure*}[ht]
    \centering
    \setlength{\abovecaptionskip}{0pt}
    \setlength{\belowcaptionskip}{10pt}
    \includegraphics[width=0.95\textwidth, trim=0 5 0 5, clip]{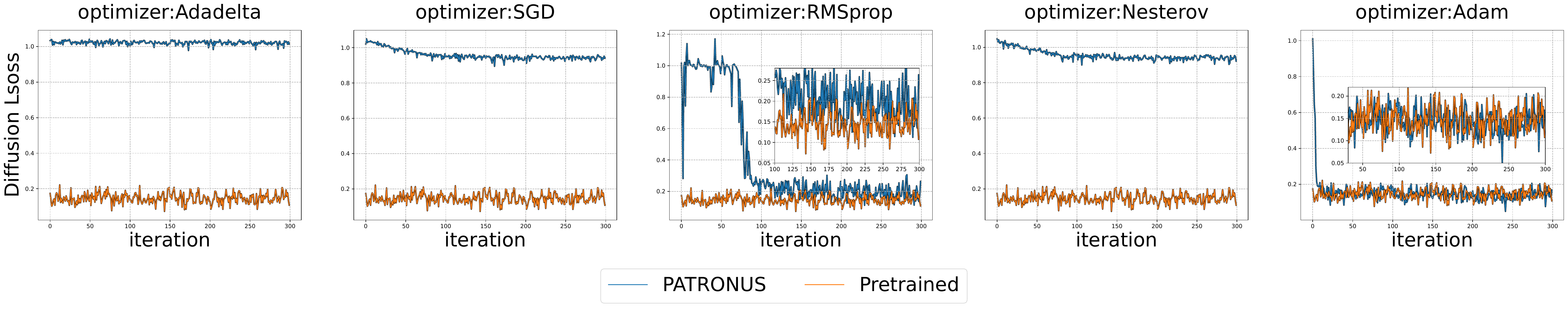}
    \caption{\sys's effectiveness of U-Net protection against different optimizers.}
    \label{fig:unet-opt}
    \vspace{-10pt}
\end{figure*}

\paragraph{Different optimizers}
As the results presented in Figure \ref{fig:op} and Appendix Table \tifs{III}, 
\sys showcases effective robustness towards different fine-tuning optimizers, including SGD, Adam, Adadelta, RMSprop, Nesterov. Specifically, SGD, Adadelta, and Nesterov optimizers fail to decrease the training loss, leading to non-functional generation models. Adam and RMSprop perform the best fine-tuning, achieving the training loss of $0.019$ and $0.022$, respectively. However, this level of training performance is still insufficient to enable unsafe image generation, given the loss of the original pre-trained model is generally around $0.003$. The robustness of \sys' defense against unseen optimizers may be attributed to the non-fine-tunable enhancement process that samples between SGD and Adam, moving the model states to a local optimum difficult for the model to escape.

\begin{figure*}[ht]
    \centering
    \setlength{\abovecaptionskip}{0pt}
    \setlength{\belowcaptionskip}{10pt}
    \includegraphics[width=0.95\textwidth, trim=0 5 0 5, clip]{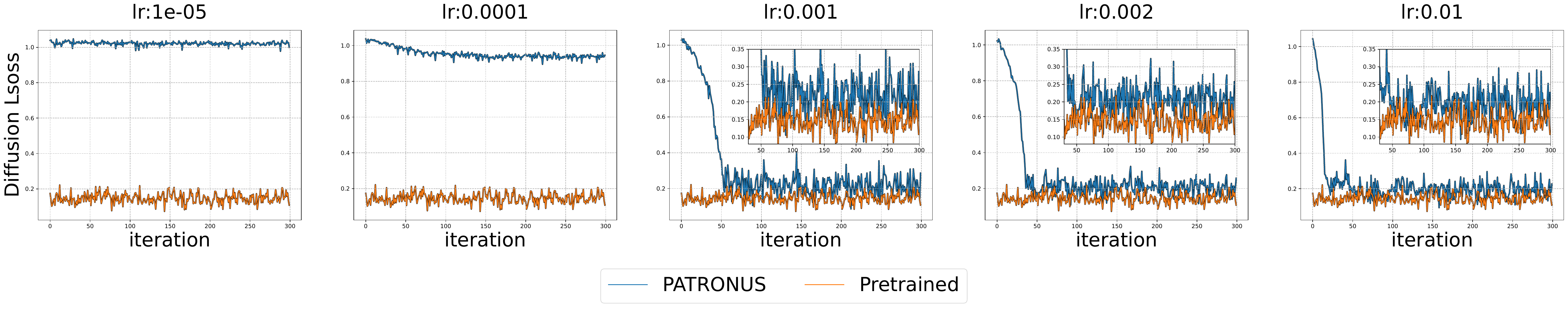}
    \caption{\sys's effectiveness of U-Net protection against different learning rates.}
    \label{fig:unet-lr}
    \vspace{-10pt}
\end{figure*}

\begin{figure*}[h]
    \centering
    \setlength{\abovecaptionskip}{0pt}
    \setlength{\belowcaptionskip}{10pt}
    \includegraphics[width=0.95\textwidth, trim=0 5 0 5, clip]{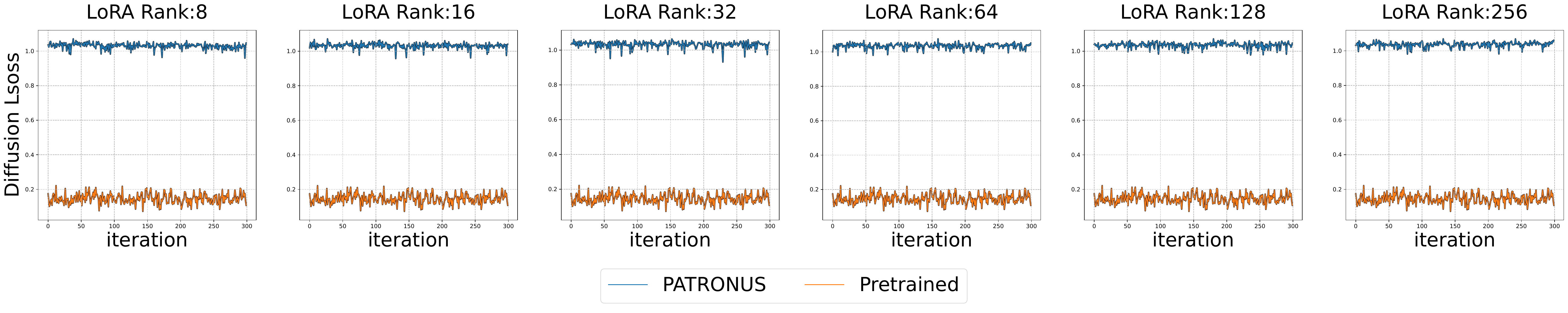}
    \caption{\sys's effectiveness of U-Net protection against different LoRA ranks}
    \label{fig:unet-rank}
    \vspace{-10pt}
\end{figure*}

\begin{figure*}[t]
\centering
    \setlength{\abovecaptionskip}{0pt}
    \setlength{\belowcaptionskip}{-10pt}
    \includegraphics[width=0.95\textwidth, trim=0 5 0 5, clip]{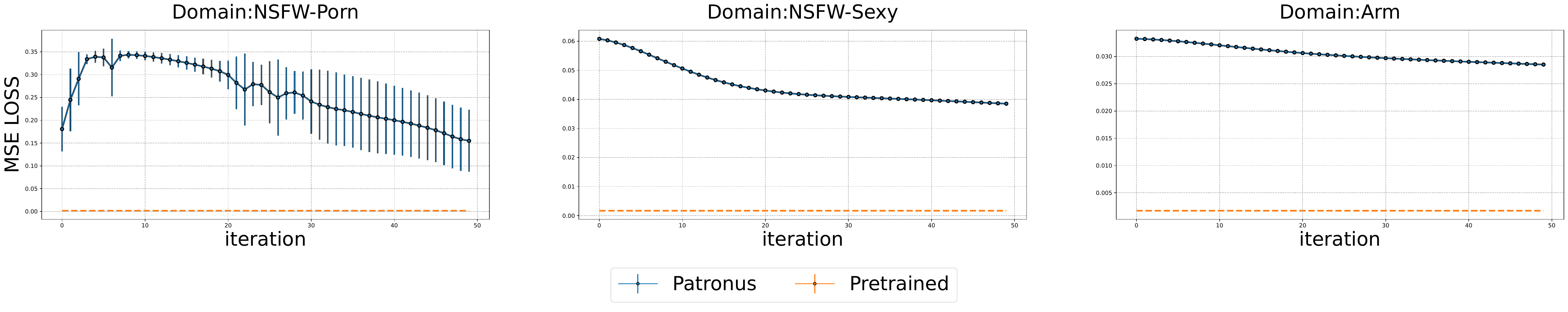}
    \caption{\sys's effectiveness of decoder protection against different unsafe categories.}
    \label{fig:domain}
\end{figure*}

\paragraph{Different learning rates}
As the results presented in Figure \ref{fig:lr} and Appendix Table \tifs{IV}, 
\sys showcases effective robustness towards different learning rate, including $1e-05$, $5e-05$, $1e-04$, $1e-03$, $2e-03$. We can see from the results that $10^{-5}$, $5\times10^{-5}$, and $10^{-4}$ lead to a slow convergence since they are too small. In contrast, $10^{-3}$ and $2\times10^{-3}$ produce a rapid convergence. However, they converge to $0.046$, $0.039$, respectively, which are far larger than the loss that allows available unsafe generation (\textit{e.g.}, typical loss of the original pre-trained model is around $0.003$). Under such suitable learning rates, \sys still resists fine-tuning, indicating its robustness against different learning rates. This could be attributed to our inclusion of varying learning rates in the fine-tuning simulation.

\paragraph{Different batch sizes}
As the results shown in Figure~\ref{fig:bs} and Appendix Table \tifs{V}, 
\sys is able to resist fine-tuning under these five batch size settings. In every case, the model fails to produce unsafe generations, as the loss remains consistently above $0.104$, which is far more than $0.003$, the typical loss of the original pre-trained model. These experiments underscore \sys's robustness against different batch sizes.

\paragraph{Different fine-tune sizes}
As the results presented in Figure \ref{fig:fs} and Appendix Table \tifs{VI}, 
\sys showcases effective robustness towards different fine-tune sizes, including $100$, $200$, $500$, $1,000$, $2,000$. We can see that as much as $2,000$ samples still cannot compromise \sys, resulting in the loss value of $0.094$, which implies that \sys can introduce great difficulty to the adversary (given that only $200$ samples are enough to corrupt ESD and SafeGen, shown in Figure \ref{fig:overall-ft}) and Figure \ref{fig:ft}. Further, we improve the fine-tuning iteration based on this setting to perform the stress test and the results are presented in Section~\ref{sec:attack-buget}.

\textbf{Adaptive Fine-Tuning Attacks towards Aligned Diffusion:}
This part assumes an adaptive adversary who knows that \sys creates a non-fine-tunable aligned diffusion module and utilizes the NSFW-\textit{prompt} image-caption dataset to implement more aggressive fine-tuning processes on the U-Net. To assess \sys's performance in the worst case, we assume the adversary has already succeeded in compromising the conditional decoder, leaving only the U-Net module to be attacked, \textit{i.e.}, we denote the T2I model with the original decoder and the defended U-Net as the subject under attack.

We assume the adversary utilizes $3000$ unsafe image-caption pairs to implement the aggressive fine-tuning processes on the U-Net. We evaluate the robustness of \sys against different fine-tuning strategies, including optimizers and learning rates, as shown in Figure \ref{fig:unet-lr} and Figure \ref{fig:unet-opt}. For the learning rates like $1e-5,1e-4$, \sys leads to the loss remaining nearly unchanged. The bigger learning rates like $0.001, 0.002, 0.01$ allow the loss to drop quickly, they converge at a larger value, leaving the model unable to generate unsafe content. We find it is also the case for RMSprop and Adam optimizers. As for other optimizers, SGD, Adadelta, Nesterov fail to decrease the training loss. We also assess \sys's effectiveness in defending LoRA (Low-Rank Adaptation \cite{hu2021lora}), a popular fine-tuning strategy in the T2I field that introduces two new low-rank parameter matrices for fine-tuning. We test different rank values to validate the robustness of \sys, as shown in Figure \ref{fig:unet-rank}.

\paragraph{Different learning rates}
As the results presented in Figure \ref{fig:unet-lr}, \sys showcases effective robustness towards different learning rate, including $1e-05$, $1e-04$, $1e-03$, $2e-03$, $1e-2$. In all cases, the loss remains higher than that of the original pre-trained model, further validating the effectiveness and robustness of \sys.

\paragraph{Different optimizers}
As the results presented in Figure \ref{fig:unet-opt}, \sys showcases effective robustness towards different fine-tuning optimizers, including SGD, Adam, Adadelta, RMSprop, Nesterov, similar to the experimental results observed for the decoder. Specifically, the loss of SGD, Adadelta, and Nesterov are significantly higher than that of the original pre-trained model, while the loss for Adam and RMSprop, although close to the original pre-trained model, are still slightly higher, indicating \sys's effectiveness in resisting fine-tuning across different optimizers.

\paragraph{Different LoRA ranks}
As the results presented in Figure \ref{fig:unet-rank}, \sys showcases effective robustness towards different LoRA Rank, including $8$, $16$, $32$, $64$, $128$, $256$. Notably, the training loss remains consistently above 1.0 for all ranks, which is significantly higher than the loss of the original pre-trained model, further validating the effectiveness and robustness of \sys in resisting unsafe generation even under different LoRA rank settings.

\subsection{Applicability for Various Unsafe Categories}
Given that existing works~\cite{li2024safegen} are often confined to the pornography category, we take a step further and evaluate the application potential of \sys against different unsafe categories. We experiment on NSFW-\textit{sexy} and the weapon dataset~\cite{3clase_dataset} as the image datasets to implement \sys and follow the similar method of I2P to build unsafe prompt sets for evaluating. Here, we consider an adaptive adversary as illustrated in Section~\ref{sec:adaptive}. We present the adversary's fine-tuning results in Figure \ref{fig:domain} and Appendix Table \tifs{VII}.
As we can see, \sys also showcases the desired rejection of unsafe content and resistance to malicious fine-tuning.

\section{Conclusion}\label{sec:conclusion}
In this paper, we introduce an innovative defense \sys for pre-trained T2I models, which includes an inseparable moderator and a non-fine-tunable safety mechanism. \sys resolves the drawbacks of existing defenses that fail to remain effective in white-box scenarios. Our experiments validate the efficacy of \sys in refusing unsafe prompting and resisting malicious fine-tuning as well as its intact benign performance.

\bibliographystyle{IEEEtran}
\bibliography{ref}

\clearpage
\appendix
\subsection{Mixed Sampling Strategy} 
\label{sec:mss}
To improve \sys's robustness against different fine-tuning processes, we propose a mixed sampling strategy for the inner loop. Specifically, we construct a bag of fine-tuning strategies containing various optimizers, learning rates, batch sizes, fine-tune sizes, and iteration numbers. Each time, we sample a fine-tuning strategy for the inner loop.

The focus of the bag of fine-tuning strategies is on the selection of the optimizer. To ensure efficiency and effectiveness, we include two optimizers in the inner optimization, \textit{i.e.}, SGD and Adam. These two optimizers have complementary dynamic characteristics, \textit{i.e.}, SGD is better at escaping local optima, while Adam is better at escaping saddle points. By resisting these two optimizers in the outer optimization, we can move our defense model to a state that is difficult to escape and performs poorly on unsafe data.

For other super-parameters like learning rates and batch sizes, we include all commonly used ranges. For fine-tuning size and iteration number, we sample from an excessive range for what is normally required for fine-tuning.
We refer to the Appendix \ref{sec:insta} for
detailed instantiations of non-fine-tunable decoders and diffusion modules.

\subsection{Instantiations of Non-fine-tunability Enhancement}
\label{sec:insta}
\textbf{Instantiate Non-fine-tunable Decoder:} For the non-fine-tunability enhancement of the decoder $\mathcal{M}_{dec}$, we designate the conditional decoder as the starting point. $\mathbbm{X}_{tune}$ and $\mathbbm{X}_{eval}$ are unsafe image sets. 
The fine-tuning-resistance loss is as follows:
\begin{equation}
    \begin{aligned}
    \label{frl1}\mathcal{L}_\mathrm{ftr}=\sum_{x_i \in\mathbbm{X}_{eval}}
\mathcal{L}_\mathrm{VGG}\left(\mathcal{M}_{dec}\left(x_i\right),\mathbf{0}\right) + \\\sum_{f_i \in\mathbbm{F}_{eval}}\mathcal{L}_\mathrm{VGG}\left(\mathcal{M}_{dec}\left(f_i\right),\mathbf{0}\right),
    \end{aligned}
\end{equation}

$\mathbbm{F}_{eval}$ is $\mathbbm{X}_{eval}$'s corresponding feature set obtained using the same method described in Section V-B2 and used for feature calibration. Optimized with this loss, the decoder learns to decode the unsafe features to smoothed zero vectors after being maliciously fine-tuned.

\textbf{Instantiate Non-fine-tunable Diffusion:} 
For the non-fine-tunability enhancement of the diffusion, we designate our aligned U-Net, which is fine-tuned to consistently predict the noise in unsafe images as zero, as the starting point. $\mathbbm{X}_{tune}$ and $\mathbbm{X}_{eval}$ are unsafe image-caption sets. Consider the U-Net, parameterized by $\theta$ (noted as 
$\epsilon_\theta$). $\epsilon_\theta$ predicts the noises added into the images.   
The fine-tuning-resistance loss is as follows:
\begin{equation}
    \begin{aligned}
    \label{frl2}
    \mathcal{L}_\mathrm{ftr}=\ &\sum_{x_i \in\mathbbm{X}_{eval}}\mathcal{L}\left(\epsilon_\theta\left(\hat{x_i}, c_i,t\right),\mathbf{0}\right)
    \end{aligned}
\end{equation}
where $c_i$ is $x_i$'s corresponding conditioning vector output by the text encoder.
Notably, the starting point we chose here is our own aligned model, though theoretically, our non-fine-tunability enhancement method can be compatible with all alignment techniques, such as SafeGen, SLD, and ESD.

\subsection{Adaptive Weighting}
\label{sec:appx-aw}
In practice, we find it difficult to assign appropriate $\gamma, \lambda$. Therefore, we refer to the Multiple Gradient Descent Algorithm (MGDA), a Multi-task learning technique
to optimize a set of (possibly conflicting) objectives.
For tasks $i = 1..k$ with respective losses $\mathcal{L}_i$, it calculates the gradient (separated from the gradients used by the optimizer) for each single task $\nabla \mathcal{L}_i$ and finds the weighting coefficients $\alpha_1..\alpha_k$ that minimize the sum
\begin{equation}
\min_{\alpha_1, \ldots, \alpha_k} \left\{ \left\| \sum_{i=1}^k \alpha_i \nabla \mathcal{L}_i \right\|_2^2 \ \middle| \ \sum_{i=1}^k \alpha_i = 1, \alpha_i \geq 0 \ \forall i \right\}.
\end{equation}

In each iteration of non-fine-tunability enhancement, we obtain $\mathcal{L}_\mathrm{ftr}$ and $\mathcal{L}_\mathrm{bpp}$, then we calculate $\gamma$ and $\lambda$ to strike a balance between $\mathcal{L}_\mathrm{ftr}$ and $\mathcal{L}_\mathrm{bpp}$, ensuring that the two tasks \textit{i.e.}, the non-fine-tunable enhancement and the benign performance preservation are simultaneously optimized (or at least not degraded).

\begin{table*}[h]\centering
\resizebox{\linewidth}{!}{
\begin{threeparttable}
\setlength{\abovecaptionskip}{0pt}%
\setlength{\belowcaptionskip}{0pt}%
\caption{Effectiveness of \sys against different optimizers.}

\setlength{\tabcolsep}{5mm}{
\begin{tabular}{l|llllll}
\toprule
\multirow{2}{*}{Optimizer}        & \multicolumn{6}{c}{Loss in the Unsafe Domain}                                          \\
                                  & Iteration 0 & Iteration 10 & Iteration 20 & Iteration 30 & Iteration 40 & Iteration 50 \\ \midrule
\multicolumn{1}{l|}{Adade}                  &   0.1267  $\pm$ 1.0e-4        &          0.1182 $\pm$ 5.5e-4    &   0.1089 $\pm$ 5.9e-4           &  0.1010 $\pm$ 9.7e-4            &      0.0939 $\pm$ 1.2e-3        &  0.0875 $\pm$ 9.5e-4            \\ 
Adam                                           &  0.4093 $\pm$ 6.2e-3           &          0.0978 $\pm$ 5.72e-3    &  0.0577 $\pm$ 4.2e-4            &  0.0503 $\pm$ 2.2e-3            &      0.0324 $\pm$ 4.3e-3        &    0.0189 $\pm$ 3.7e-3          \\ 
Nes                                                &  0.2388 $\pm$ 6.2e-2           &         0.3403 $\pm$ 7.3e-3     &   0.3036 $\pm$ 1.6e-2           &  0.2315 $\pm$ 8.1e-2            &   0.1847 $\pm$ 7.5e-2           &  0.1332 $\pm$ 6.9e-2            \\
RMS                                                  &  0.7481 $\pm$ 2.2e-1           &           0.0526 $\pm$ 7.0e-3   &  0.0374 $\pm$ 1.6e-2            & 0.0250 $\pm$ 5.1e-3             &      0.0195 $\pm$ 3.0e-3        &  0.0216 $\pm$ 1.2e-2            \\
SGD                                                   & 0.1807 $\pm$ 4.9e-2            &        0.3427 $\pm$ 8.6e-3      &   0.3075 $\pm$ 2.3e-2           &  0.2541 $\pm$ 5.3e-2            &   0.2033 $\pm$ 7.7e-2           &  0.1549 $\pm$ 6.8e-2            \\
\bottomrule
\end{tabular}

}

\label{tab:optimizer}
\end{threeparttable}}
\end{table*}

\begin{table*}[h]
\centering
\resizebox{\linewidth}{!}{
\begin{threeparttable}
\setlength{\abovecaptionskip}{0pt}%
\setlength{\belowcaptionskip}{0pt}%
\caption{Effectiveness of \sys against different learning rate.}

\setlength{\tabcolsep}{5mm}{
\begin{tabular}{l|llllll}
\toprule
\multirow{2}{*}{Learning Rate}        & \multicolumn{6}{c}{Loss in the Unsafe Domain}                                          \\
                         & Iteration 0 & Iteration 10 & Iteration 20 & Iteration 30 & Iteration 40 & Iteration 50 \\ \midrule
\multicolumn{1}{l|}{0.001}     &   0.3379  $\pm$ 2.3e-2        &          0.2074 $\pm$ 3.0e-2    &   0.0660 $\pm$ 9.7e-3           &  0.0557 $\pm$ 5.8e-3            &      0.0500 $\pm$ 2.4e-3       &  0.0459 $\pm$ 3.0e-3            \\ 
0.00005                    &  0.1418 $\pm$ 2.0e-2           &          0.1835 $\pm$ 8.1e-2    &  0.1427 $\pm$ 9.4e-2            &  0.0611 $\pm$ 2.7e-2            &      0.0463 $\pm$ 5.9e-3       &    0.0422 $\pm$ 3.1e-3         \\ 
0.0001                      &  0.1807 $\pm$ 4.9e-2           &         0.3427 $\pm$ 8.6e-3     &   0.3075 $\pm$ 2.3e-2           &  0.2541 $\pm$ 5.3e-2            &   0.2033 $\pm$ 7.7e-2           &  0.1549$\pm$ 6.8e-2            \\
0.002                   &  0.3517 $\pm$ 1.2e-2           &           0.1202 $\pm$ 3.3e-2   &  0.0681 $\pm$ 1.4e-2            & 0.0499 $\pm$ 1.5e-3             &      0.0449 $\pm$ 3.0e-3        &  0.0394 $\pm$ 2.7e-3            \\
0.00001                        & 0.1272 $\pm$ 2.4e-4            &        0.1174 $\pm$ 2.2e-3      &   0.0974 $\pm$ 2.9e-3         &  0.0818 $\pm$ 3.3e-3            &  0.0695  $\pm$  8.9e-4          &  0.0616 $\pm$ 3.9e-4            \\
\bottomrule
\end{tabular}

}

\label{tab:lr}
\end{threeparttable}}
\end{table*}

\begin{table*}[h]\centering
\resizebox{\linewidth}{!}{
\begin{threeparttable}
\setlength{\abovecaptionskip}{0pt}%
\setlength{\belowcaptionskip}{0pt}%
\caption{Effectiveness of \sys against different batch size.}

\setlength{\tabcolsep}{5mm}{
\begin{tabular}{l|llllll}
\toprule
\multirow{2}{*}{Batch Size}        & \multicolumn{6}{c}{Loss in the Unsafe Domain}                                          \\
  & Iteration 0 & Iteration 10 & Iteration 20 & Iteration 30 & Iteration 40 & Iteration 50 \\ \midrule
\multicolumn{1}{l|}{5}    &   0.2021  $\pm$ 5.9e-2        &          0.3409 $\pm$ 1.0e-2    &   0.3047 $\pm$ 2.0e-2           &  0.2027 $\pm$ 9.7e-2            &      0.1662 $\pm$ 8.4e-2       &  0.1250 $\pm$ 8.1e-2            \\ 
10                            &  0.2418 $\pm$ 5.7e-2           &          0.3216 $\pm$ 4.3e-2    &  0.3101 $\pm$ 5.3e-3            &  0.2606 $\pm$ 1.4e-2            &      0.1813 $\pm$ 6.8e-2       &    0.1180 $\pm$ 5.8e-2         \\ 
15                      &  0.2021 $\pm$ 5.9e-2           &         0.3409 $\pm$ 1.0e-2     &   0.3047 $\pm$ 2.0e-2           &  0.2027 $\pm$ 9.7e-2            &   0.1662 $\pm$ 8.4e-2           &  0.1250 $\pm$ 8.1e-2            \\
20                                 &  0.1807 $\pm$ 4.9e-2           &           0.3427 $\pm$ 8.6e-3   &  0.3075 $\pm$ 2.3e-2            & 0.2541 $\pm$ 5.3e-2             &      0.2033 $\pm$ 7.7e-2        &  0.1549 $\pm$ 6.8e-2            \\
30                                 & 0.1854 $\pm$ 5.6e-2            &        0.3400 $\pm$ 7.4e-3      &   0.3055 $\pm$ 1.9e-2           &  0.2257 $\pm$ 7.0e-2            &    0.1923 $\pm$ 6.1e-2            &   0.1038 $\pm$ 1.2e-2             \\
\bottomrule
\end{tabular}

}

\label{tab:batchsize}
\end{threeparttable}}
\end{table*}

\begin{table*}[h]
\centering
\resizebox{\linewidth}{!}{
\begin{threeparttable}
\setlength{\abovecaptionskip}{0pt}%
\setlength{\belowcaptionskip}{0pt}%
\caption{Effectiveness of \sys against different Finetune number.}

\setlength{\tabcolsep}{5mm}{
\begin{tabular}{l|llllll}
\toprule
\multirow{2}{*}{Finetune number}        & \multicolumn{6}{c}{Loss in the Unsafe Domain}                                          \\
                          & Iteration 0 & Iteration 10 & Iteration 20 & Iteration 30 & Iteration 40 & Iteration 50 \\ \midrule
\multicolumn{1}{l|}{100}       &   0.1873  $\pm$ 4.1e-2        &          0.3361 $\pm$ 6.3e-3    &   0.2936 $\pm$ 1.4e-2           &  0.1934 $\pm$ 7.9e-2            &      0.1136 $\pm$ 6.2e-2       &  0.0727 $\pm$ 2.5e-2            \\ 
200                     &  0.1599 $\pm$ 3.5e-2           &          0.3299 $\pm$ 3.4e-3    &  0.2751 $\pm$ 3.7e-2            &  0.1424 $\pm$ 8.0e-2            &      0.0999 $\pm$ 7.8e-2       &    0.0836 $\pm$ 6.7e-2         \\ 
500                    &  0.1807 $\pm$ 4.9e-2           &         0.3427 $\pm$ 8.6e-3     &   0.3075 $\pm$ 2.3e-2           &  0.2541 $\pm$ 5.3e-2            &   0.2033 $\pm$ 7.7e-2           &  0.1549 $\pm$ 6.8e-2            \\
1000                    &  0.2263 $\pm$ 2.5e-2           &           0.3009 $\pm$ 8.1e-2   &  0.2619 $\pm$ 1.0e-1             &      0.2001 $\pm$ 1.0e-1        &  0.1443 $\pm$ 8.3e-2   & 0.0959 $\pm$ 7.0e-2        \\
2000              & 0.1806 $\pm$ 5.3e-2            &        0.2742 $\pm$ 9.8e-2      &   0.2056 $\pm$ 1.3e-1           &  0.1425 $\pm$ 1.2e-1            &  0.1211  $\pm$  9.8e-2          & 0.0942  $\pm$  7.7e-2           \\
\bottomrule
\end{tabular}

}

\label{tab:ft_number}
\end{threeparttable}}
\end{table*}

\begin{table*}[h]
\centering
\resizebox{\linewidth}{!}{
\begin{threeparttable}
\setlength{\abovecaptionskip}{0pt}%
\setlength{\belowcaptionskip}{0pt}%
\caption{Effectiveness of \sys against different (potentially) unsafe topics.}

\setlength{\tabcolsep}{5mm}{
\begin{tabular}{l|llllll}
\toprule
\multirow{2}{*}{Domain}        & \multicolumn{6}{c}{Loss in the Unsafe Domain}                                          \\
                                          & Iteration 0 & Iteration 10 & Iteration 20 & Iteration 30 & Iteration 40 & Iteration 50 \\ \midrule
\multicolumn{1}{l|}{NSFW-porn}         &   0.1807 $\pm$  4.9e-2       &          0.3427 $\pm$ 8.6e-3      &     0.3075 $\pm$ 2.3e-2      & 0.02541  $\pm$ 5.3e-2   & 0.2033 $\pm$ 7.7e-2          &  0.1549 $\pm$ 6.8e-2       \\ 
NSFW-sexy                                     &   0.0608 $\pm$ 3.4e-5          &      0.0518 $\pm$ 3.0e-4          &    0.0435 $\pm$ 8.9e-5       &  0.0410 $\pm$ 2.0e-5           &            0.0398 $\pm$ 7.8e-5 & 0.0385 $\pm$ 2.8e-5 
\\ 
Weapon                                    & 0.0333 $\pm$1.9e-6            &      0.0322 $\pm$ 2.8e-5        &  0.0308 $\pm$ 2.8e-5            &  0.0298 $\pm$ 1.9e-5            &   0.0291 $\pm$ 1.9e-5           &   0.0285 $\pm$ 8.4e-6           \\ \bottomrule
\end{tabular}

}

\label{tab:domain}
\end{threeparttable}}
\end{table*}

\end{document}